\documentclass[iop,apj,twocolumn,numberedappendix,twocolappendix]{openjournal}

\usepackage{xcolor}
\usepackage{newtxtext,newtxmath}
\usepackage{textgreek}
\usepackage[utf8]{inputenc}
\usepackage[english]{babel}
\usepackage{hyperref}
\hypersetup{breaklinks,colorlinks,citecolor=blue,urlcolor=cyan}
\usepackage{color,colortbl}
\usepackage{tensind}
\tensordelimiter{?}
\DeclareGraphicsExtensions{.bmp,.png,.jpg,.pdf}
\usepackage{verbatim}
\usepackage[normalem]{ulem}
\usepackage{orcidlink}
\usepackage{soul}
\graphicspath{ {./}{./figs/} }

\newcommand{\Msun}{\,{\mathrm{M}_\odot}}
\newcommand{\Omegatid}{\Omega_{\mathrm{tid}}}
\newcommand{\feh}{\mathrm{[Fe/H]}}

\begin{document}
\title{\vspace{-6mm} Catalog of Mock Stellar Streams in Milky Way-like Galaxies\vspace{-14mm}}

\author{Colin Holm-Hansen$^\star$\orcidlink{0009-0002-1128-2341}}
\author{Yingtian Chen\orcidlink{0000-0002-5970-2563}}
\author{Oleg Y. Gnedin\orcidlink{0000-0001-9852-9954}}
\affiliation{Department of Astronomy, University of Michigan, Ann Arbor, MI 48109, USA}

\thanks{$^\star$E-mail: \href{mailto:cphh@umich.edu}{cphh@umich.edu}}

\begin{abstract}
Dynamically cold stellar streams from tidally dissolved globular clusters (GCs) serve as excellent tools to measure the Galactic mass distribution and show promise to probe the nature of dark matter. For successful application of these tools to observations, it is essential to have models of stellar stream properties on the Galactic scale. To this end we produce a mock catalog of stellar streams in four Milky Way-mass galaxies from cosmological simulations. We build the catalog with three main components: a model for the formation and disruption of globular clusters based on cosmological simulations, time-dependent potentials constructed with basis function expansions for integrating stream orbits, and an improved particle-spray algorithm for efficient generation of stellar streams. We generate mock photometry for \textit{Gaia}, LSST, and \textit{Roman}, and find that the latter two surveys will increase the number of observable stars in GC stellar streams by several orders of magnitude, with LSST alone being able to observe 1000 or more stars in 40-100 streams beyond the stellar disk. In addition, we find that the observable widths and lengths of mock streams as a function of galactocentric radius are well described by power-laws with slopes of $-1.8$ and $-2$ for streams beyond 10~kpc. Our full catalog, containing stream populations across four different galaxy realizations, is publicly available and can be used to study stream population statistics and to calibrate models which use stellar streams to understand our Galaxy.
\end{abstract}


\maketitle

\section{Introduction}
\label{sec:intro}

Globular clusters (GCs) are some of the best available tools for understanding the assembly history and evolution of the Milky Way galaxy (MW). Since most clusters are observed to have an age $\sim 10$ Gyr \citep{Krauss2003, Dotter2011}, they have been orbiting our Galaxy for at least a few Gyr if accreted onto the MW (ex-situ formed clusters) or longer if born in the central part of the MW (in-situ formed clusters). While GC stars can become unbound through internal dynamics such as two-body relaxation \citep{Spitzer1987, Meylan1997} and ejection from black hole interactions \citep{GielesGnedin2023, Roberts2025MNRASParticleSpray}, the dominate mechanism is through tidal heating \citep{Gnedin1997, Gnedin1999ApJ, BaumgardtMakino2003MNRAS}, which can cause enough mass-loss to form an unbound stellar stream. Recently, the \textit{Gaia} mission \citep{Gaia2018, Gaia2023} has significantly expanded our knowledge of MW streams \citep{BonacaPriceWhelan2025}, of which $\sim$80 are thought to originate from GCs \citep{Mateu2023}. Since the velocity dispersion of unbound GC stars is smaller than the typical orbital velocity around the MW, GC streams are dynamically cold, retaining signatures of their tidal histories for up to several Gyr. This makes them ideal for studying a wide range of gravitational physics within the MW itself, both on small and large scales.

One of the most exciting applications which shows promise to be better understood through stellar streams is constraining the mass distribution within the Galaxy \citep{Johnston1999ApJPotentialConstrain, Helmi1999Nature, Koposov2010ApJ, Sanders2013MNRAS, Bonaca2018ApJ, Reino2021MNRAS, Ibata2024ApJStreamFinderII, Nibauer2025ApJ}. This is typically done by assuming an analytic form of the gravitational potential and then varying the potential's free parameters and/or symmetry constraints to identify which model can best recreate a given stream or select group of streams. In addition, \cite{Nibauer2025ApJ} showed that the GD-1 stream \citep{Grillmair2006ApJ} can be used to constrain the potential by direct differentiation of the stream track in phase space, without first having to assume an analytic form for the potential. Stream modeling can also be used to infer properties of satellite galaxies interacting with the MW. For example, \cite{Shipp2021LMC} used measurements from the Dark Energy Survey \citep[DES;][]{DES2016}, the Southern Stellar Stream Spectroscopic Survey \citep[$S^5$;][]{Li2019S5}, and \textit{Gaia} to estimate the mass of the Large Magellanic Cloud (LMC) by modeling its perturbations on different stellar streams, obtaining a value of $1.88^{+0.35}_{-0.40} \times10^{11}\Msun$. \cite{Vasiliev2021SgrStream} performed a similar type of analysis to recreate \textit{Gaia} measurements of the Sagittarius stream \citep{Ibata2001ApJSgrStream} in the presence of the LMC, computing a MW virial mass of $(9.0\pm{1.3})\times10^{11}\Msun$.

In addition to modeling the global potential of the MW and its satellites, streams are also subject to perturbations from dark matter subhalos \citep[e.g., ][]{Johnston2002ApJ, Ibata2002MNRASsubhalos, Carlberg2012ApJ, Erkal2016MNRASDMSubhalos, Banik2018JCAP} which can in principle be used to constrain the subhalo mass function. However, the density of stars within streams may also exhibit variations caused by baryonic perturbations from substructure in the Galactic potential, such as a tilted stellar disk \citep{Nibauer2024ApJDiskTilt}, giant molecular clouds \citep{Amorisco2016MNRAS}, and the Galactic bar \citep{Pearson2017GalacticBar, Bonaca2020ApJPal5Model}. Epicyclic overdensities can also appear as density variations in the stream track without a perturbation \cite{Kupper2008MNRAS, Kupper2010MNRAS}. Disentangling these additional perturbations from those potentially caused by dark matter is key in the hope of using stellar streams to constrain the dark matter subhalo mass function. Detecting streams at large galactocentric radii will be extremely important for this purpose, where interactions with baryonic components of the potential are less likely. A key effort in this direction is to model streams beyond the stellar disk ($r_{\rm gal} \sim 15 \rm~kpc$), both in terms of physical properties and prospects for observing them.

In order to model these GC streams far out from the Galactic center, one must first have an understanding of how the GC population itself is evolving over time for a given model Galaxy, as well as a method for generating streams from the clusters based on their mass-loss rates along a given orbit. This was done in \cite{Panithanpaisal2025arXiv}, which applied the GC formation and evolution models from \cite{Grudic2023MNRAS} and \cite{Rodriguez2023MNRAS} to study two GC streams evolved in a MW-like Galaxy from the FIRE-2 cosmological simulation suite \citep{Wetzel2016ApJ, Hopkins2018MNRAS, Wetzel2023ApJS}. Looking at the real MW GC population, \cite{Ferrone2023} modeled the theoretical extra-tidal features of MW GCs for which 6D phase space coordinates are known and found coherent stream-like structures should be present as far out as $300$ kpc from the Galactic center.

In studies such as these which model many stream realizations in a realistic Galactic potential, an approximation to direct N-body is needed to speed up the computation time. For this purpose, many works utilize the particle-spray method, which releases stream particles near the Lagrange points at frequent timesteps with small offsets in their positions and velocities \citep{Varghese2011MNRASParticleSpray, Lane2012MNRASParticleSpray, Kupper2012MNRASParticleSpray, Bonaca2014ApJParticleSpray, Gibbons2014MNRASParticleSpray, Fardal2015, Roberts2025MNRASParticleSpray}. Integrating the orbits of these particles to the present day produces an elongated stream profile, generally consistent with the observations, providing a reasonably accurate method which is significantly faster than full N-body modeling methods. Recently, \cite{YBChen2024ParticleSpray} developed an improved particle-spray algorithm where the phase-space offsets of released particles follow a multivariate Gaussian distribution. The new method matches the morphology and action space distribution of streams from N-body simulations with an error of only 10\%. This method allows computationally efficient modeling of Galactic stream populations with greater detail than previously possible. 

Previous work has shown that due to the slow rate at which streams phase-mix with the Galactic stellar population, there is likely to be a large population of GC streams that remain distinguishable from field stars \citep{Sandford2017MNRAS, Carlberg2018ApJ}, with many of them still undetected by the current observational facilities. In particular, \cite{Pearson2024} modeled the evolution of streams in a static MW potential using the \cite{Fardal2015} particle-spray method and found that there may be up to $\sim 1000$ coherent GC stellar streams in the MW. Of these, more than $100$ coherent GC stellar streams are predicted to be beyond $15 ~\rm kpc$ from the Galactic center, which would be observable by the LSST survey on the Vera C. Rubin Observatory with 10-year stacking data \citep[$g < 27.5$;][]{LSSTCollab2009}.
 
In this paper, we build upon the work of \cite{Pearson2024} and generate mock GC stream populations in four simulated MW-like galaxies with realistic time-evolving potentials. We use the \cite{YBChen2024ParticleSpray} particle-spray algorithm to integrate the orbits of unbound stars for several Gyr. We analyze overall properties of the final stream populations and study how facilities such as LSST and the future Nancy Grace Roman Space Telescope \citep[\textit{Roman}; ][]{Spergel2013Roman, Spergel2015Roman} will expand our ability to detect Galactic GC streams. We also make our stream catalog publicly available to be used as a tool for analyzing expected trends in the stream populations, and for calibration of models which attempt to understand the Galaxy through the use of GC streams.

The paper is organized as follows: in Section~\ref{sec:model} we describe components of the model used to generate the catalog and introduce the data structure of the catalog itself. In Section~\ref{sec:Results} we describe properties of the streams in the catalog and evaluate photometric predictions for observing the streams. We discuss these results in further detail and provide more possible use cases of the catalog in Section~\ref{sec:Discussion}. We conclude with a summary of our results in Section~\ref{sec:conclusion}.

\section{Model Setup}
\label{sec:model}

In this section, we describe the details of the modeling used to generate our mock stream catalog. The key components are the initialization of the GC population, dynamical evolution of the GCs, and the generation of streams via the particle-spray method. 

\subsection{Globular Cluster Evolution}
\label{sec:gc_mass}

The modeling of GC evolution requires a detailed understanding of not only the birth of clusters, both in-situ and ex-situ, but also their mass loss over time through stellar evolution and tidal disruption. For these purposes, we employ the GC evolution model by \cite{Chen2022MNRASModel1}, which is further developed in \cite{YBChen2024MNRAS}. The model post-processes snapshots of a selected cosmological simulation and creates catalogs of the GC population, keeping track of cluster formation and evolution. The model produces key observables including mass, metallically, positions, and velocities of individual GCs. The model has been applied to a wide range of galaxies in cosmological simulations and can successfully match the observed properties of the Galactic GC system.

A full description of the most recent version of the model is given in \cite{YBChen2024MNRAS}. The three main features are: cluster formation, particle assignment for orbit tracking, and mass loss through tidal disruption along the GC's orbit. GC formation is triggered when the host galaxy mass accretion rate exceeds a specified threshold, which is an adjustable parameter of the model. The total mass of the cluster population formed at a given snapshot epoch is proportional to the cold gas mass of the parent galaxy, following \cite{Kravtsov2005}. Given this population, the individual masses of GCs are sampled from a \cite{Schechter1976} initial cluster mass function with a lower bound of $10^{4}\Msun$ and an exponential cutoff of $10^{7}\Msun$. To obtain kinematic information for the clusters at this epoch, the model assigns clusters to collisionless particles in the background cosmological simulation and uses their positions and velocities. The mass evolution takes into account both stellar evolution and tidal disruption along the cluster orbit. Once a cluster is initialized with mass $M_0$, the model treats the mass-loss due to stellar evolution as instantaneous:
\begin{equation}
    M_{\rm i} = \mu_{\rm sev} M_0
\end{equation}
where the remaining fraction $\mu_{\rm sev}$ is taken to be $0.55$ \citep{GielesGnedin2023}. This assumption is justified because the timescale for stellar evolution of high mass stars (tens of Myr) is much shorter than the typical age of GCs. Here $M_{\rm i}$ represents the initial mass of the cluster after stellar evolution but before any tidal disruption.

The stream is populated by stars unbound from the cluster by tidal forces. The most recent prescription for tidal mass-loss outlined in \cite{YBChen2024MNRAS} is given by:
\begin{equation} 
    \label{masslossrate}
    \frac{dM}{dt} = -2\times 10^4 \frac{M_{\odot}}{\text{Gyr}}\left(\frac{M_{\rm i}}{2\times10^5 M_{\odot}} \right)^{\frac{1}{3}}\left(\frac{M(t)}{M_{\rm i}} \right)^{-\frac{1}{3}}\left(\frac{\Omegatid(t)}{150\,\text{Gyr}^{-1}}\right)
\end{equation}
where $\Omegatid(t)$ is the effective tidal frequency, approximated via the first and third eigenvalues ($\lambda_1, \lambda_3$) of the tidal tensor of the host potential:
\begin{equation}
    \Omegatid(t)^2 = \lambda_1 - \lambda_3
    \label{eq:omegatid}
\end{equation}
where the tidal tensor is composed of the second derivatives of the Galactic potential $\Phi$. In Cartesian coordinates, it takes the form:
\begin{equation}
  \mathrm{T}_{ij} = -\frac{\partial^2\Phi}{\partial x^i\partial x^j}
\end{equation}

The model calculates the tidal tensor using a second order finite difference method for snapshots where the potential for the host galaxy is provided. However, this only permits the masses of the GCs to be updated at every full snapshot. To improve the time resolution, we use a combination of the GC masses computed by the \cite{YBChen2024MNRAS} model as well as a higher-resolution calculation of the mass-loss rate outlined in Section~\ref{sec:stellar stream generation}. Our mass-loss rate model is calibrated from N-body models ran in \cite{GielesGnedin2023} and the parameters were optimized in \cite{YBChen2024MNRAS} to be able to match the MW in terms of total GC number, present day mass function, radial distribution, metallicity distribution, and velocity dispersion. While other phenomena can affect individual GC mass-loss rates that are either not incorporated or may be unresolved in Equation \ref{masslossrate} such as GC concentration \citep{BaumgardtMakino2003MNRAS}, heating from black holes \citep{GielesGnedin2023}, and tidal shocks \citep[e.g., ][]{Gnedin1999ApJ}, statistically our model matches the present day trends of the total MW GC population.

We note that by default, the GC evolution model calculates the mass of each GC at every snapshot, but only outputs the mass to the user at the present day. To address this for the purposes of this work, we make a small modification of the code to have it output the cluster masses at every snapshot. Due to this modification, we re-run the model on the four cosmological halos utilized in this work. Both re-running the model as well as utilizing a different mass-loss calculation method results in our GCs being slightly different from the GCs in the published catalogs of \citet{YBChen2024MNRAS}. These differences are explained in detail in Appendix \ref{Properties of Surviving GC Population}, but we stress that overall our GCs are still tailored to match the MW system in terms of the mass, metallicity, and radial distributions.

\subsection{Background Cosmological Simulations}
\label{sec:simulation}

To generate a realistic model for an initial GC population in MW-like galaxies, we utilize two halos from the Illustris TNG50-1 (TNG) cosmological simulation \citep{Nelson2019, Pillepich2019MNRAS}, and two halos from the Latte suite of FIRE-2 (FIRE) simulations \citep{Wetzel2016ApJ, Hopkins2018MNRAS, Wetzel2023ApJS}. The TNG galaxies have subhalo IDs 523889 and 519311 in the TNG halo catalog, and the FIRE halos are m12i and m12w. \cite{YbChen2024OJAGalaxyAssembly} identified these four halos to be a close match to the Milky Way for the purpose of applying their GC post-processing model. The criteria used for determining suitable galaxies in \cite{YbChen2024OJAGalaxyAssembly} include having a present mass and circular velocity similar to that observed for the MW, no major merger (defined as having a mass ratio greater than 1:4) in the last 10~Gyr, one major merger between 10-12~Gyr ago representing the Gaia-Sausage/Enceladus merger \citep{Belokurov2018GSE, Helmi2018GSE}, and to have formed 25-35\% of its stellar mass by 10~Gyr ago \citep{Leitner2012ApJ}. Our selected halos satisfy all these requirements and are suitable candidates to approximate the present day stream distribution of the MW.

TNG and FIRE provide individual snapshots of halo particles, separated by $0.1-1$~Gyr, which by itself is not suitable for computing the orbits of millions of stream stars for several Gyr. For the actual computation of stream orbits, we construct analytical basis function expansion (BFE) potentials based on the TNG and FIRE snapshots using the \texttt{agama} software \citep{Agama2019}. We construct the analytic time-dependent potentials in the following way: for each snapshot, we fit a BFE potential to the snapshot given the positions and masses of the stellar, gas, and dark matter particles. The dark matter particles are fit with a standard multipole expansion of spherical harmonics with $l_{\text{max}}$ taken to be $12$. The stellar and gas particles are fit with an azimuthal Fourier harmonic expansion with $m_{\text{max}} = 6$. The method for determining these parameters and further details are outlined in Appendix~\ref{ap:ap}. We construct the time-variable potentials by linearly interpolating the expansion coefficients between the snapshots. For each simulation the overall time span is taken to be $z=0.3$ to the present day (about 3.5~Gyr). For the TNG halos, this corresponds to every snapshot starting from 78 to 99 (present day), which are spaced by intervals ranging from $120-235$ Myr for TNG's cosmological parameters. For FIRE, we use the first data release snapshots with full particle data from $z=0.3$ to present, corresponding to snapshots 446, 486, 534, and 590-600. The spacing between these snapshots respectively is 987~Myr, 1140~Myr, 1280~Myr, and 2.22~Myr for the remaining ten.

For the FIRE halo m12i, we find that the principal axis of the disk tilts by $24^{\circ}$ over this time span, and therefore we modify the above prescription for this halo only and model its potential just with multipole expansion. This is because all orbits are computed in the galactocentric frame of the halo at the last snapshot, but the \texttt{agama} Fourier harmonic expansion fitting procedure requires the disk to be aligned with the $X-Y$ plane to work properly. Since multipole expansion is not as effective at modeling flattened systems, this results in our disk potential for m12i not being as accurate as in the other three halos. However, we find that after fitting one of our other galaxies with only multipoles, the median separation between our reconstructed orbit to the true orbit at the last snapshot increases by only about 2\% compared to using both BFE functions. Since this difference is small, we still proceed to use the m12i halo, but users of the catalog should be aware of this if they choose to use the m12i streams.

Our method of potential reconstruction through interpolation of BFE potentials has been shown to be an effective way for approximating the potentials of galaxies from cosmological simulations \citep[e.g., ][]{Lowing2011MNRAS, Sanders2020MNRAS, Arora2024ApJ}. In Appendix \ref{Reconstructed Orbits}, we show some example orbits of our analytic BFE potentials compared to the orbits of the GC particles in their respective cosmological simulations. In general, this prescription of potential modeling allows us to reliably recover the pericenter and apocenter values of the GC particles over our integration time. As other works have already shown, we find the method is less effective on orbits whose apocenters are extremely close to the Galactic center, $\lesssim 5$ kpc. However, these orbits are less likely to have kinematically cold streams to begin with. We also find that for some GCs with shorter periods, the reconstructed orbit becomes out of phase with the true one as small errors accumulate over many orbits. \citep{Arora2024ApJ} found specifically that pericenters and apocenters can be recovered within $10\%$ over a baseline of 4 Gyr. While they use a higher temporal cadence than what is utilized in this work, we show in Appendix \ref{Reconstructed Orbits} that our method also reasonably recovers the period, pericenter, and apocenter of most orbits over a similar integration time, which are key quantities in determining the overall stream morphology.

Instead of integrating orbits of stream particles immediately from cluster birth (or from time of accretion onto the MW), we select a lookback time to begin tracking the orbits of tidally stripped stars from each cluster. This is because phase mixing causes the orbits of evaporated stars to deviate significantly from the progenitor with time, limiting the duration of coherent streams on most orbits to a few Gyr. Moreover, \cite{YBChen2025ApJMassLossGCs} showed that the duration of most observed GC stellar streams are $\leq 1 \rm ~Gyr$ (although some streams are likely to be much older – see \citealt{Erkal2016MNRASDMSubhalos}). It is thus unnecessary to include these early-released particles in our catalog. Based on these arguments we take our initial snapshot for orbit tracking of evaporated particles to be $z = 0.3$.

We stress that the main benefit of our time-dependent potential is that it captures the mass evolution of the galaxies over the course of the GC orbits. For each halo there is significant mass growth over $3.5$ Gyr and therefore this effect is important. In principle, a time dependent potential could capture other aspects of a dynamic Galactic potential on stellar streams, such as the spiral arms and the Galactic bar. However, the cadence of the simulation snapshots is insufficient to accurately capture these effects in the smoothed BFE potential.

\subsection{Stellar Stream Generation}
\label{sec:stellar stream generation}

In this section, we describe the complete process for stream generation, given the Galactic potential and our initial GC conditions from the model of \cite{YBChen2024MNRAS}. The distinct inputs to our stream model are as follows:
\begin{itemize}
    \item The Galactic potential $\Phi(\textbf{x},t)$
    \item The GC masses at birth and at $z = 0.3$
    \item Positions and velocities of GCs at $z=0.3$
    \item The cluster ages (at present day) and metallicities
\end{itemize}

We use the Galactic potentials as described above in Sec.~\ref{sec:simulation}. Every other item is supplied from the GC formation model, which we treat as input for generating our streams. Our stream generation has three main components: calculation of the cluster mass-loss rate, selection of released stream stars, and orbit integration of the stream stars.

\subsubsection{Mass Loss Calculation}
\label{highresmlrate}

The overall mass-loss calculation is in two parts. From cluster birth until $z=0.3$, we use the mass loss given by the GC formation model. Then beginning at $z=0.3$, we numerically solve Equation~\ref{masslossrate} given the tidal tensor along the cluster orbit computed with \texttt{agama} to calculate the amount of mass lost by the cluster from $z=0.3$ until the present day. While \texttt{agama} uses an adaptive timestep for orbit integration, we interpolate the solution such that the mass history and tidal tensor components are provided at 1000 uniformly spaced intervals over the integration time, giving us both quantities with a time resolution of $\sim3.5$~Myr. Figure~\ref{fig:agamatngcomp} in Appendix~\ref{TidalFreqAppendix} shows the effective tidal frequency as a function of time at the provided resolution for an example cluster in one of the TNG halos. There are only four full TNG snapshots with potential information in our chosen time of integration, which miss significant variation of the tidal force. Consequently, the \texttt{agama} framework much more accurate for detailed calculations of the mass loss.

\subsubsection{Stream Realization}

Once calculated, the mass-loss histories are passed onto the next step, where we initialize a stellar population for each cluster and compute the number and masses of stream stars to be released at each timestep. First, individual cluster stars are sampled according to a \cite{Kroupa2001} initial mass function (IMF) with the initial cluster mass given by the GC model. For the lower and upper bounds of the IMF, we use MIST isochrones \citep{Choi2016ApJ, Dotter2016ApJ} though the \texttt{isochrones} interface \citep{Isochrones2015ascl} to calculate the theoretical lowest and highest mass stars that would still be luminous today given the cluster's age and metallicity. These values vary from cluster to cluster but are generally around $0.1\Msun$ for the lower bound and between $0.8-0.9\Msun$ for the upper bound. Stars with higher masses than this are already removed from the cluster mass accounting by stellar evolution in \S\ref{sec:gc_mass}. We then evolve the mass function by removing stars until the sum of the remaining stellar mass is equal to the cluster mass at $z=0.3$. Since models of star cluster evolution \citep{BaumgardtMakino2003MNRAS} show that low-mass stars tend to evaporate preferentially compared to high-mass stars, we remove stars probabilistically such that low-mass stars are more likely to be evaporated. Specifically, we treat the probability that any given star be removed as proportional to the inverse square root of its mass $m$:
\begin{equation}
    p_{\rm remove} \propto m^{-1/2}
\end{equation}
based on energy equipartition arguments. We continuously remove stars and re-normalize the probabilities until the sum of the remaining stellar masses is equal to the mass of the cluster at $z=0.3$. The remaining pool of stars are considered to be the bound cluster stars before generating the stream.

To further assess what effect the evaporation probability has on the stream profile, we selected a cluster and generated its stream twice, once with the adopted ejection probability $p \propto m^{-1/2}$ and once where every star is equally likely to be ejected independent of mass. After running the model, we binned the stars in the stream according to mass such that there were approximately 500 stars in each bin, and looked for variations between the two models as a function of mass. We found none significant. For example, the width of the stream, which we quantify with the standard deviation along the short axis of the stream, $\sigma_{\phi_2}$, shows little dependence on stellar mass: the best-fit slope of the linear regression was $-0.16 \pm 0.015$ for the stream with weighted ejection probability, and $-0.010\pm0.018$ for the stream with uniform ejection probability. Both of these slopes are small, and we conclude that the evaporation probability has no significant effect on the structure of the stream. We proceed to use the probabilistic ejection setup in our model to represent observations that older parts of streams have lower mass stars \citep{Balbinot2018MNRAS}.

\begin{center}
    \begin{figure}[!t]
        \centering
    	\includegraphics[width=\columnwidth]{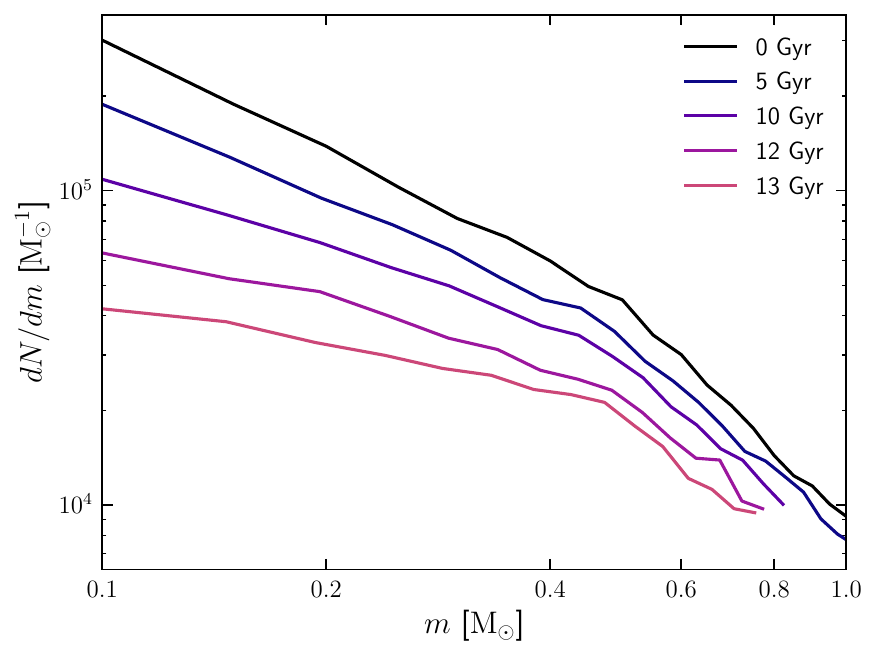}
    	\caption{Evolution of the mass function over time for a $2\times10^{4}\Msun$ cluster on a Pal 5-like orbit with stellar ejection probability $p \propto m^{-1/2}$ and assumed metallicity of $\feh = -1.56$ (the observed metallicity of Pal 5). The cluster becomes less dominated by low-mass stars gradually over time because they are more likely to be ejected than high-mass stars.}
        \label{fig:mfevolve}
    \end{figure}
\end{center}

\begin{figure*}[t]
        \centering
    	\includegraphics[width=\textwidth]{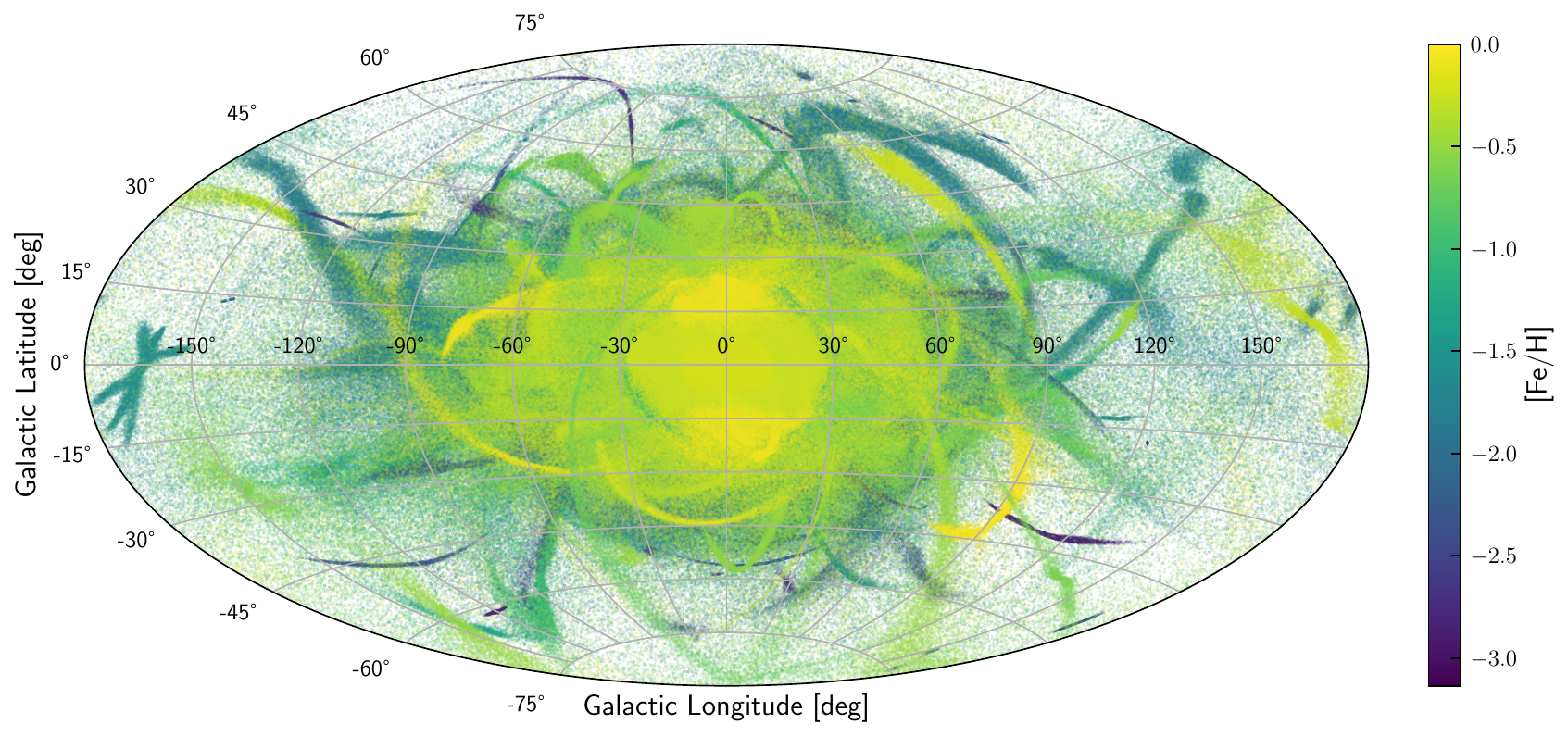}
    	\caption{All-sky map of the final collection of disrupted features for the TNG halo 523889 with $g < 25$. Color indicates the metallicity of the original star cluster. Positions and apparent magnitudes of stars are calculated with the standard \texttt{astropy} convention where the Sun is placed at $(-8 \rm ~kpc, 0 ~kpc, 20.8 ~pc)$ in our galactocentric reference frame.}
        \label{fig:allskyplot}
\end{figure*}

Analytically, the number of stars unbound from the cluster $N_{\rm stars}$ will depend on the mass-loss rate along the cluster's orbit and the remaining mass function of stars in the cluster. This can be expressed as:
\begin{equation} 
\label{eqn:nstars}
    N_{\rm stars}(t) = \int^t \frac{\dot{M}(t')}{m_{\rm av}(t')}~dt'
\end{equation} 
where $\dot{M}(t)$ is the mass-loss rate along the progenitor orbit and $m_{\rm av}(t)$ quantifies the average mass of ejected stars at time t. In terms of the ejection probability for a given star $p(m,t)$ (time-dependent because the normalization factor decreases as the number of stars in the cluster decreases) and the mass function of the cluster $\phi(m,t)$, $m_{\rm av}$ is equal to
\begin{equation} 
\label{eqn:xi}
    m_{\rm av}(t) = \frac{\int p(m,t)\, m\, \phi(m,t)\, dm}{\int p(m,t)\, \phi(m,t)\, dm}
\end{equation} 

Figure~\ref{fig:mfevolve} shows how the mass function of an example cluster in our model changes over time given the ejection probability of the individual stars. We take a cluster with initial mass $2\times10^4\Msun$, a Kroupa IMF with lower and upper bounds of $0.1-1.1\Msun$, similar apocenter and pericenter parameters as Pal 5 \citep{Odenkirchen2001ApJ}, and integrate its orbit in a static version of one of our TNG potentials (523889) at the last snapshot for $13$~Gyr. We remove stars from the cluster given its mass-loss rate from Equation \ref{masslossrate}. At each age plotted, we update the upper mass bound based on the current location of the main sequence turn off, which varies slightly with cluster metallicity. For this figure we adopt a metallicity similar to Pal 5 of $\feh = -1.56$ \citep{Koch2017Pal5FeH}, which corresponds to a maximum mass of $0.8\Msun$ at present with the MIST isochrones. For the approximate metallicity range relevant to GCs, roughly between $-2.5$ and $-0.5$~dex, the maximum mass at 13~Gyr ranges from $0.8 - 0.88\Msun$. After several Gyr, the slope of the low-mass end becomes less steep as low-mass stars are preferentially stripped from the cluster. This illustrates that streams realized with individual stars cannot be modeled by simply applying a zero-age IMF, because the cluster evolution significantly changes the mass distribution of the stream, especially for streams which are nearly or completely disrupted.

We assign stream particles numerically according to Equations~\ref{eqn:nstars} and \ref{eqn:xi}. At each timestep, we take the mass-loss rate of the cluster and calculate how much mass needs to be removed. Then, we remove 2 stars at a time (one for the leading tidal tail and one for the trailing, without replacement) from the bound stellar pool using the same method as described above, until the total mass released at a given timestep equals the appropriate amount according to the cluster mass-loss rate. This is repeated until the present day, giving us $N_{\rm stars}(t)$.

\subsubsection{Orbit Integration}

Given the number of stars released at each timestep, we integrate the stream orbits from when the stars are released (up to $\sim3.5$~Gyr ago) until the present day. Using the \cite{YBChen2024ParticleSpray} particle-spray method, stars are released near the Lagrange points, with their exact phase-space coordinates sampled from a multivariate gaussian distribution. This phase space distribution utilized by the \cite{YBChen2024ParticleSpray} algorithm was tested in static, axisymmetric Milky Way potentials, which are inherently less complex than the potentials used in this work, which are both time-evolving and non-axisymmetric. In Appendix \ref{NbodyComp}, we do a comparison with one stream from our catalog generated with full N-body, and verify that the \cite{YBChen2024ParticleSpray} shows excellent agreement with the N-body realization even in our time-evolving, non-axisymmetric regime. 

Each orbit is integrated using our particle-spray model in the combined potential of the galaxy and the cluster's self-gravity, which is modeled as a Plummer sphere \citep{Plummer1911}. We continuously decrease the mass of the Plummer sphere according to the evolving mass of the cluster. We set the Plummer scale radius of each cluster to $4~\mathrm{pc}$. While this value should decrease as the cluster loses mass, its exact value will not effect the orbit of escaped stars if the cluster's tidal radius is much greater than the scale radius. For our clusters that do not fully disrupt, we find this to be true in almost all cases; for fully disrupted clusters, we find this to be true down until the cluster mass is in the range $10^2-10^3\Msun$. As such, we keep the scale radius fixed for simplicity. Stars are released until the present day or until the cluster is fully disrupted, whichever occurs first. Once every stellar orbit has been completed we are left with the final stream resolved with individual stars at the present day. Repeating this process for each GC gives us the final stream catalog for each halo.

\begin{figure*}[!t]
    \centering
    \includegraphics[width=\textwidth]{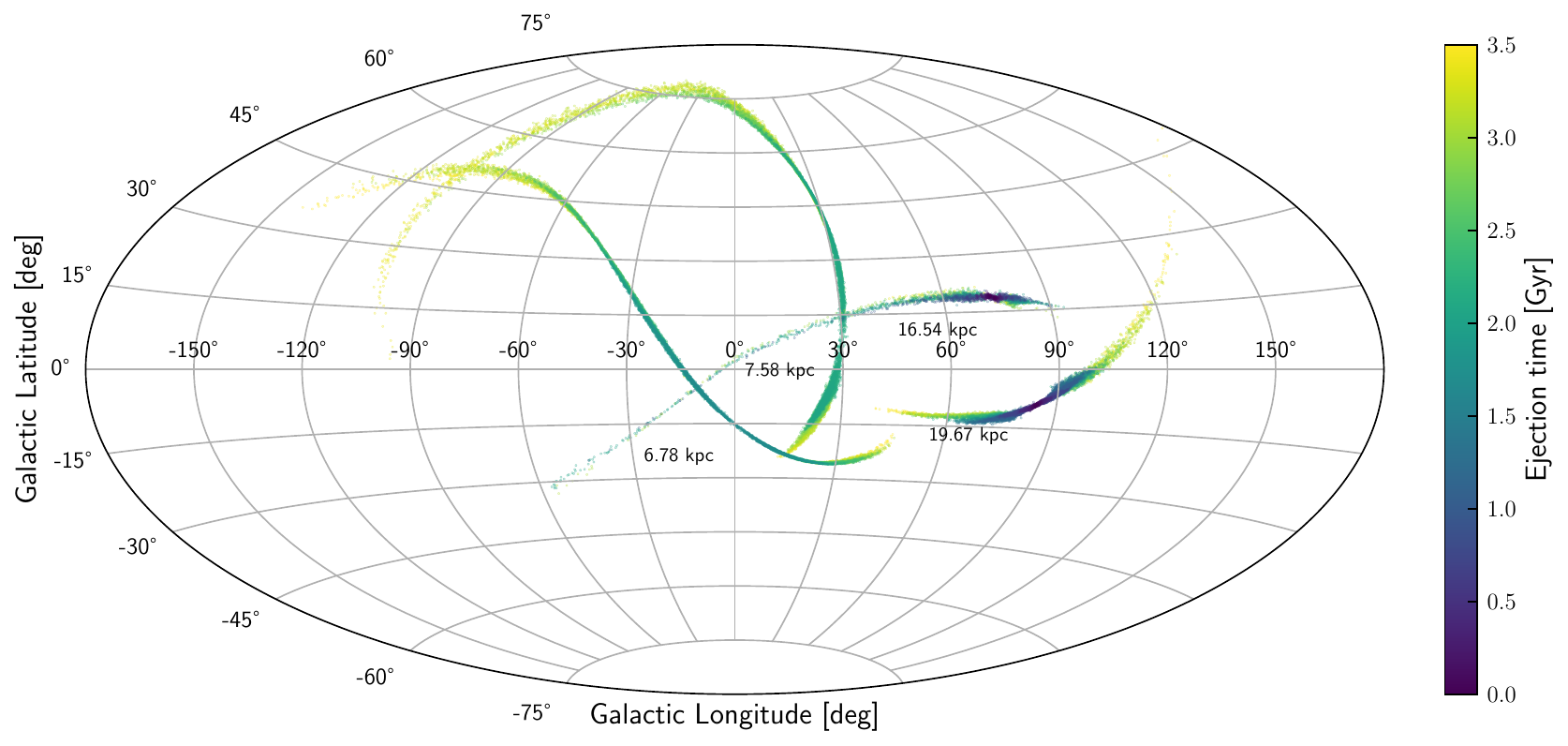}
    \caption{Four example streams with stars color-coded by their ejection lookback time. Each stream's present day galactocentric radius is shown down and to the left of its progenitor location, as indicated by region where the ejection time approaches either 0~Gyr or the time of disruption.}
    \label{fig:allsky_examplestreams}
\end{figure*}

\begin{figure*}[!t]
    \centering
    \includegraphics[width=\textwidth]{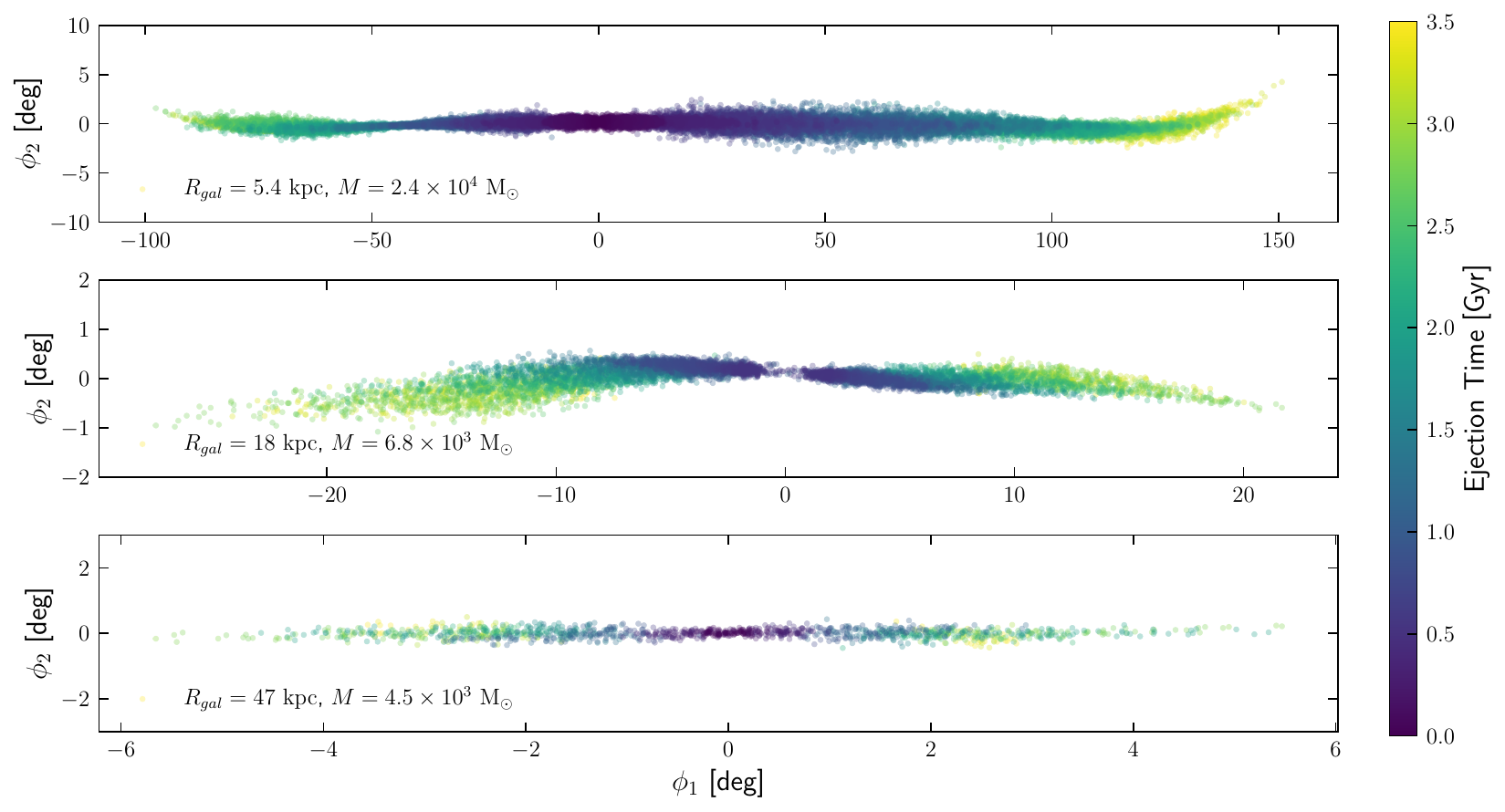}
    \caption{Examples of streams from the catalog at a range of present day galactocentric radii with a photometric cut $g < 25$. The top stream subtends the longest extent on the sky at nearly $250^{\circ}$ and its close proximity to the Galactic center has heated it and caused the stream to extend in $\phi_2$. The middle stream progenitor fully disrupted approximately $470 \rm ~Myr$ ago, causing a gap along the stream profile. The final stream at $47$~kpc from the Galactic center is dynamically the coldest, making it the shortest and thinnest of the three shown.}
    \label{fig:along_examplestreams}
\end{figure*}

\section{Results}
\label{sec:Results}

In this section we calculate overall statistics for our catalog of disrupted GCs and make photometric predictions for present and upcoming missions. For plots we adopt the \texttt{astropy} convention and assume the Solar System to have galactocentric coordinates of $(-8~\mathrm{kpc}, 0, 20.8~\mathrm{pc})$ in a right-handed system. Unless otherwise stated, we display only stars with an apparent magnitude $g < 25$, the limit of a single visit for LSST ~\citep{LSSTCollab2009}. All our photometry calculations are done with MIST Isochrones \citep{Choi2016ApJ, Dotter2016ApJ}, and stream properties are calculated only using stars below our photometric magnitude cuts. We also note that for our photometric predictions, we do not include extinction effects.

\subsection{Catalog Overview and Statistics}

Our model produces a full catalog of the tidal remnants of GCs that had non-zero mass at a lookback time $\sim$3.5~Gyr. The catalog provides data for each stream and their progenitor GCs in two data files. The first file contains masses, positions, velocities, energies, and ejection lookback times in the galactocentric frame for each star at the present time, for each stream. The second file contains information on the progenitor GCs, including their mass history during the period of stream generation, the positions and velocities of their orbits, the tidal tensor along their orbits over the period of stream generation, the present-day age, and $\feh$ for each cluster. Our repository on Github\footnote{\href{https://github.com/cholm-hansen/StreamCatalogs}{https://github.com/cholm-hansen/StreamCatalogs}} contains a description of the catalog, provides instructions for downloading it, and contains an example notebook showing how to calculate properties of the streams.

In total, our four halos (523889, 519311, m12i, m12w) contain 287, 401, 391, and 392 streams respectively. Table \ref{tab:halosummary} shows for each halo the total number of streams, the total mass of stars in streams, as well as the total number of stars in streams. Figure~\ref{fig:allskyplot} shows every stream for one of the TNG halos (523889) plotted on the sky with our fiducial magnitude cut. Each remnant is colored by its metallicity, which is the same value for each star of a given cluster. There is rich variation in the structure of tidal features across the entire population. Toward the Galactic center, there is a spherical grouping of stars from multiple clusters similar in morphology to the Galactic bulge \citep{Wegg2013MNRAS}. These are predominantly clusters that have very small pericenters, where the tidal field is too strong to allow long, thin streams. At higher Galactic latitudes, there is an approximately uniform distribution of stars which have deviated significantly from their progenitor, identifiable as members of the stellar halo. Finally, there are many long and thin features which are clearly identifiable as stellar streams.

\begin{figure*}[!t]
    \centering
    \includegraphics[width=\textwidth]{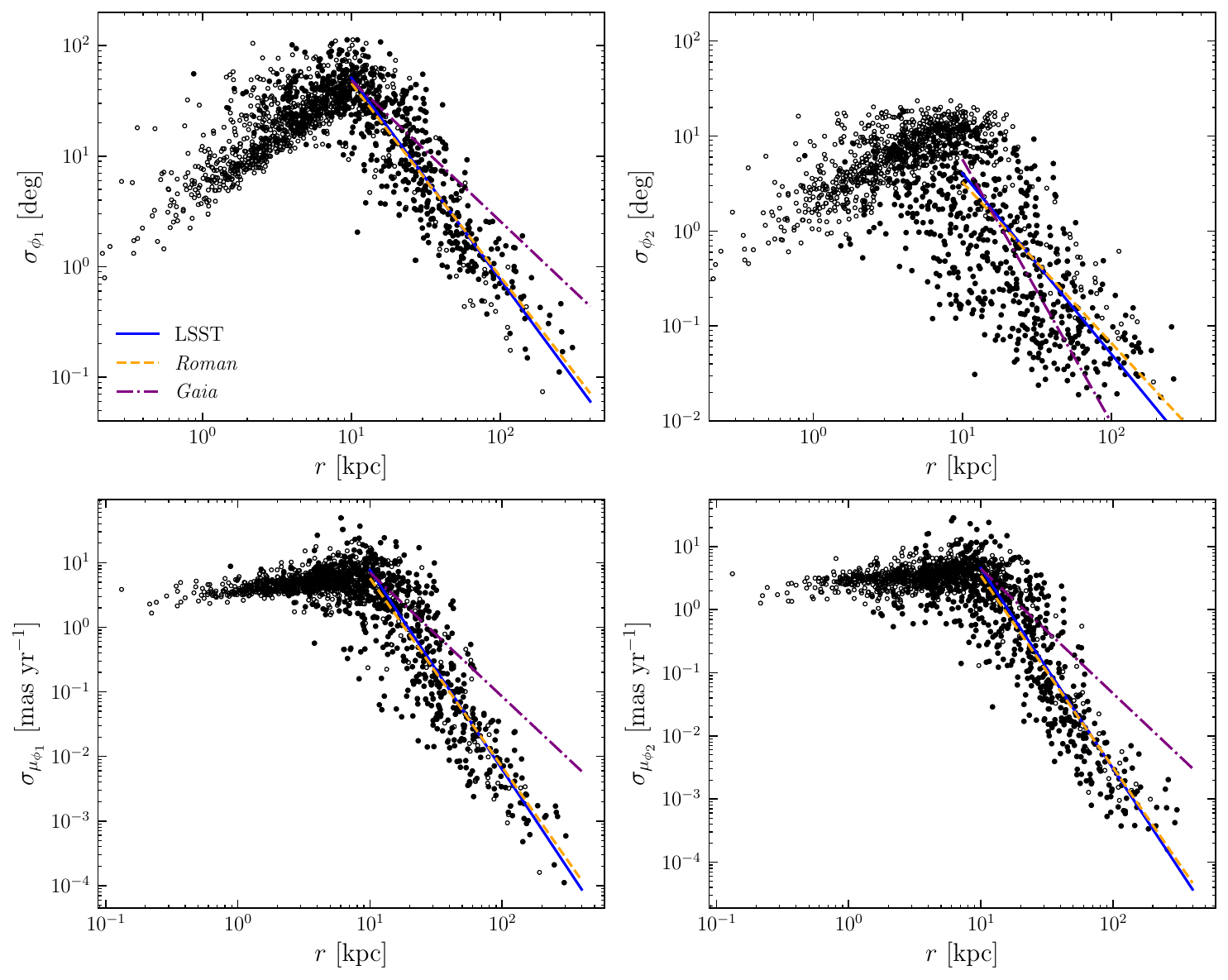}
    \caption{Variation of each stream in $(\phi_1, \phi_2)$ coordinates across all four halos as a function of the galactocentric radius of the progenitor orbit for stars with $g < 25$. The top row shows the dispersion for each stream in both $\phi_1$ and $\phi_2$, where as the bottom row shows the dispersion in the proper motions of each coordinate. Filled circles show thin streams, while open circles show wide streams. Streams closer to the Galactic center are dynamically hotter than streams further away and therefore exhibit a larger variation. Lines show best-fit power law relations at $r > 10$~kpc for the three magnitude cuts: LSST (solid line), \textit{Roman} (dot line), and \textit{Gaia} (dot-dashed line).}
    \label{fig:phi1phi2plots}
\end{figure*}

Not every cluster's tidal feature has a morphology that would be traditionally identified by an observer as a stream. We investigated distributions of the width and axis ratios of all features in the catalog to see if there was an obvious way to group them into 'classical' streams with more characteristic thin tidal tails and wider streams. For this purpose, for each GC we first rotate its observed tidal debris on the sky into a frame aligned with the two principle components of the escaped stars, denoted as $(\phi_1, \phi_2)$ in angular sky coordinates. This is done by first constructing a rotation matrix $R$ via the eigenvalues and eigenvectors of the covariance matrix of star positions $(x, y, z)$. We then rotate the stars into this new frame $(x', y', z')$, where for a thin stream, $x'$ is orthogonal to the long axis of the stream, $y'$ is aligned with the long axis, and $z'$ is orthogonal to the short axis. $\phi_1$ is then the longitude angular coordinate (length) and $\phi_2$ is the latitude angular coordinate (width) in the stream-aligned frame. We find that inspection of the stream population in $(\phi_1, \phi_2)$ coordinates does not reveal evidence of two distinct populations. In light of this, and the fact that there is no consensus definition of a stream in the literature, we use the term 'stream' for all objects in the catalog but we separate them into \textit{thin} and \textit{wide} streams below.

The average combined mass of all streams in each halo is $2\times10^7\Msun$, ranging from $1.6\times10^7\Msun$ (m12w) to $2.5\times10^7\Msun$ (519311). The average total mass of stars in features within $3$ kpc of the Galactic center, roughly corresponding to the radius of the Galactic bulge, is $1.3\times10^7\Msun$, ranging from $0.93\times10^7\Msun$ (m12w) to $1.8\times10^7\Msun$ (519311).
Figure~\ref{fig:allsky_examplestreams} shows four example thin streams from halo 523889, which span a range of present-day progenitor galactocentric radii $5-20$~kpc. Stars are colored by how long ago they were released from the cluster. A different set of streams spanning $5-50$~kpc is shown in Figure~\ref{fig:along_examplestreams}. For each stream in this Figure, the progenitor location is placed at $\phi_1 = 0^{\circ}$. The innermost stream, at a present day galactocentric radius of $5.4$ kpc, is the widest and has noticeable width variations along its track. The progenitor GC of the stream in the middle panel is fully disrupted at the present day, causing a gap to have formed in its center. There is also a slight asymmetry in the tidal tails caused by the projection into the $\phi_1-\phi_2$ frame. The outermost stream, at $47$ kpc, is the shortest, thinnest, and least massive, being subjected to the weakest tidal field out of the three. Collectively, these Figures illustrate the wide range of features and morphologies present in the overall MW stream population.

\begin{table}[]
    \renewcommand\arraystretch{1.2}
    \centering
    \caption{Overall statistics for each halo including the number of streams, total number of stars in streams, and total mass of streams for each halo.}
    \begin{tabular}{l|c|c|c}
       Halo ID & N$_{\rm streams}$  & N$_{\rm stars} ~[10^7]$  & M$_{\rm streams} ~[10^7 \Msun]$
       \\[1mm] \hline
       523889   & $287$ & $6.1$ & $2.0$ \\
       519311 & $401$ & $7.8$ & $2.5$\\
       m12i & $391$ & $5.4$ & $1.7$ \\
       m12w & $392$ & $5.1$ & $1.6$ \\
    \end{tabular}
    \label{tab:halosummary}
\end{table}

\begin{figure}[!t]
    \centering
    \includegraphics[width=\columnwidth]{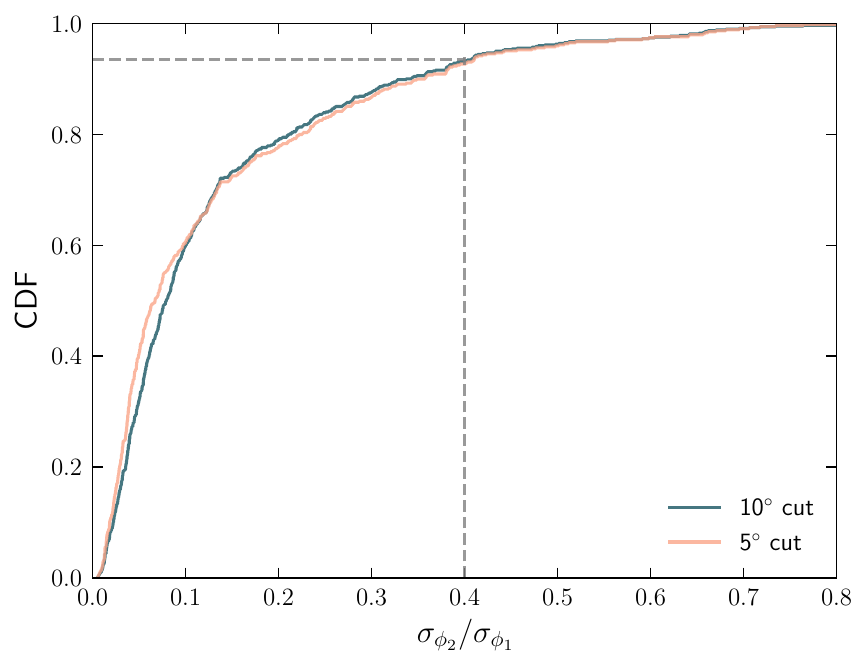}
    \label{fig:phi1phi2cdf}
    \caption{Cumulative distribution of the stream width-to-length ratio with two upper limits on the width: $\sigma_{\phi_2} < 5^{\circ}$ and $\sigma_{\phi_2} < 10^{\circ}$. We adopt $\sigma_{\phi_2}/\sigma_{\phi_1} = 0.4$ based on the flattening of the CDF at that location to differentiate thin streams from wide streams.}
    \label{fig:thin}
\end{figure}

\subsection{Thin vs. Wide Streams}

Figure~\ref{fig:phi1phi2plots} shows how standard deviation in $\phi_1$, $\phi_2$, $\mu_{\phi_1}$\footnote{Here and in the rest of the paper, we use $\mu_{\phi_1}$ as shorthand for $\mu_{\phi_1}\cos{\phi_2}$.}, and $\mu_{\phi_2}$ vary as a function of distance to the Galactic center, as would be observed from Earth. Here, $\sigma_{\phi_1}$ characterizes the length of the stream, and $\sigma_{\phi_2}$ characterizes its width. For the proper motion dispersions and stream length, we calculate each quantity by rotating into the stream frame and computing the standard deviation of $\mu_{\phi_1}, \mu_{\phi_2}, $ and $\phi_1$. For the width of the stream, we first use \texttt{agama} to fit a spline to the stream track in $(\phi_1, \phi_2)$ coordinates and calculate the standard deviation from the fitted track. These statistics are calculated only using the stars in each stream that are below our fiducial magnitude cut, $g < 25$.

We distinguish here between what we consider wide streams (open circles) and thin streams (filled circles). While wide streams often in the literature refer to dwarf galaxy streams, here we take wide streams to be GC streams which have undergone significant heating due to their orbit. These two groups are classified in the following way: first, we rotate each stream into galactocentric $(\phi_1, \phi_2)$ coordinates. This is done to emphasize the heating caused by an orbit being close to the Galactic center, and to de-emphasize observation effects, such as a stream with a blob-like morphology which appears thin due to its distance from the Earth. Any stream with $\sigma_{\phi_2}$ exceeding $10^{\circ}$ is considered a wide stream, which captures features that are too thick to be traditionally classified as a thin stream. Then we make a cumulative distribution of the ratio of a stream's width to its length, $\sigma_{\phi_2}/\sigma_{\phi_1}$, in Figure~\ref{fig:thin}. Based on the flattening of the CDF gradient, we classify thin streams as having $\sigma_{\phi_2}/\sigma_{\phi_1} < 0.4$, corresponding to axis ratio greater than 2.5:1. We also repeated this procedure with an initial cut of $\sigma_{\phi_s}$ exceeding $5^{\circ}$ to check if our choice of initial cut significantly affects the CDF. As can be seen from Figure~\ref{fig:thin}, there is a minimal difference between an initial cut of 5 or 10 degrees; for our classification we use 10 degrees.

\begin{table}[]
    \renewcommand\arraystretch{1.2}
    \centering
    \caption{Power Law fits for the streams as a function of galactocentric distance}
    \begin{tabular}{l|c|c|c|c}
       Instrument & $\sigma_{\phi_1}$  & $\sigma_{\phi_2}$  & $\sigma_{\mu_{\phi_1}}$ & $\sigma_{\mu_{\phi_2}}$
       \\[1mm] \hline
       LSST   & $-1.8\pm{0.3}$ & $-1.9\pm{0.6}$ & $-3.0\pm{0.5}$ & $-3.2\pm{0.5}$ \\
       \it{Roman} & $-1.8\pm{0.3}$ & $-1.7\pm{0.6}$ & $-2.9\pm{0.5}$ & $-3.0\pm{0.5}$ \\
       \it{Gaia} & $-1.3\pm{0.4}$ & $-2.8\pm{0.9}$ & $-1.9\pm{0.7}$ & $-2.0\pm{0.7}$ \\
    \end{tabular}
    \label{tab:powerlawfits}
\end{table}

Beyond the inner galaxy, these relations all appear to follow a power law, but with significant intrinsic scatter. For each quantity $X$, we use maximum likelihood estimation to calculate the best-fitting power law relation of the form $X \propto r^{\alpha}$ for all points beyond $r > 10$~kpc, because at smaller radii the relation visually appears to break down. This is influenced by the solar radius being at $\approx 8$ kpc, but we also find that the relation breaks around the same radius when the quantities in Figure~\ref{fig:phi1phi2plots} are calculated as seen from the Galactic center. To estimate the uncertainties for each slope, we resample our data $1000$ times and use the bootstrapping technique. For the standard deviation in $\phi_1$ and $\phi_2$ (in the top row of Figure~\ref{fig:phi1phi2plots}) the best-fit slopes are $\alpha= -1.8\pm0.3$ and $\alpha = -1.9\pm 0.6$. For the dispersion in proper motions along $\phi_1$ and $\phi_2$ (the bottom row of the figure), the slopes are $\alpha=-3.0\pm0.5$ and $\alpha=-3.2\pm 0.5$. 
 
We also recompute what the slopes would be using magnitude cuts for \textit{Roman} (dashed line) and \textit{Gaia} (dot-dashed line). Details of how the cuts are made are in the next section (\ref{sec:photpreds}). We show the full list of best-fit slopes for each instrument in Table \ref{tab:powerlawfits}. Overall, both LSST and \textit{Roman} are consistent with each other, with the prediction from \textit{Gaia} varying significantly in comparison. This can be explained from the fact that the \textit{Gaia} samples are mostly incomplete at large galactocentric radii, with only a few observable stars in streams beyond $30$ kpc. As a result, the width relation from \textit{Gaia} is underestimated at large galactocentric radii, whereas the other three quantities are overestimated. Most significant of these results is that width estimates for GC streams with member stars confirmed with \textit{Gaia} may be underestimated. In addition to the scatter already present in the width distribution, it is likely that variation in GC concentration (which is not captured in this work) would add additional scatter.

\begin{figure}[!t]
    \centering
    \includegraphics[width=\columnwidth]{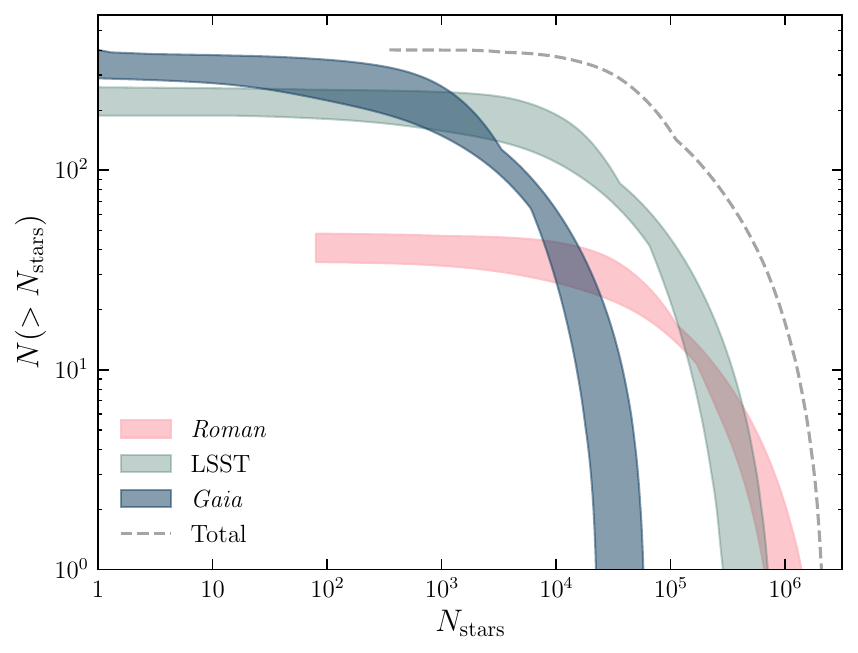}
    \caption{Cumulative distribution of the number of streams with $N_{\rm stars}$ or more observable stars according to photometric completeness limits for \textit{Gaia} ($G < 19$), LSST ($g < 25$), and \textit{Roman} ($z < 27.5$). Based on their reported magnitude limits, LSST and \textit{Roman} will both drastically expand our ability to identify MW stellar streams. Dashed line shows the total number of streams.}
    \label{fig:gaiakde}
\end{figure}

\subsection{Stream Photometric Predictions}
\label{sec:photpreds}

In this section, we investigate the observability of our streams with \textit{Gaia}, LSST, and \textit{Roman}. For each stream, we calculate the apparent magnitude of each star in the $g$ band (peaked at 0.48$\mu$m), $G$ band (0.67$\mu$m), and $z$ band (0.91$\mu$m) using MIST. We then calculate the number of stars in each stream which would be observable with each instrument. For \textit{Gaia} and LSST, we take this to be $G < 19$ and $g < 25$, respectively. While \texttt{isochrones} does not have \textit{Roman} filters available for photometric calculations, the \textit{Roman} web documentation\footnote{\href{https://roman-docs.stsci.edu/roman-instruments-home/wfi-imaging-mode-user-guide/introduction-to-the-wfi/wfi-quick-reference}{https://roman-docs.stsci.edu/roman-instruments-home/wfi-imaging-mode-user-guide/introduction-to-the-wfi/wfi-quick-reference}} reports that the F087 filter's most similar ground equivalent is the $z$ band, and that for point sources the expected depth for the F087 filter is an AB-magnitude of 27.63 in a one-hour pointing. Based on this we take $z < 27.5$ as the threshold to be considered observable by \textit{Roman}. With each observability threshold set, we use gaussian kernel density estimation (KDE) to obtain probability density functions (PDFs) for the number of observable stars in each stream, for each instrument.

We integrate our PDFs to get cumulative distributions of the number of streams that contain $N_{\rm stars}$ observable stars for a given instrument, for a given halo: $N(> N_{\rm stars})$. This result is shown in Figure~\ref{fig:gaiakde}. For each instrument, the region plotted encapsulates all four individual distributions for each halo. The uncertainty arises from the differing number of total streams in each halo. For \textit{Roman} and LSST, we account for sky coverage by scaling each CDF according to what percent of the sky the instrument is expected to cover: $65\%$ for LSST, and $12\%$ for \textit{Roman}'s High-Latitude Wide-Area Survey (although in the future \textit{Roman} is expected to observe other portions of the sky as well). We also plot the theoretical limiting distribution if every stream star in the catalog were observable. The figure shows how impactful LSST and \textit{Roman} will be in identifying new streams. For example, \textit{Gaia} is only able to observe at least $10^4$ stars in about $60$ streams, but LSST would be able to observe the same number of stars in hundreds of streams. \textit{Roman} would be able to detect at least 1000 stars in $30-40$ streams, which corresponds to essentially every stream in its available region of sky across all four halos. 

\begin{figure}[!t]
    \centering
    \includegraphics[width=\columnwidth]{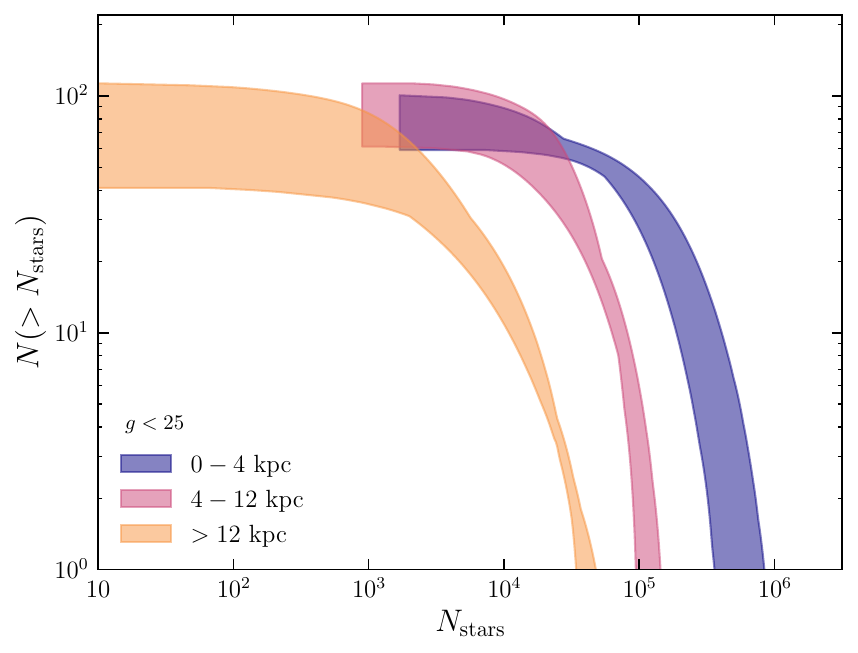}
    \caption{Number of streams with $N_{\rm stars}$ or more observable stars by LSST binned by their present-day galactocentric radii. Each shaded region encapsulates the range between all four halos. For example, the number of streams beyond 12 kpc with 100 or more observable stars ranges from 40 to over 100 depending on the halo. The uncertainty arises primarily from the varying number of streams in each halo. This predicts that LSST has the potential to greatly expand our census of known streams beyond the stellar disk.}
    \label{fig:gaia_radii}
\end{figure}

In Figure~\ref{fig:gaia_radii} we divide the streams into three approximately equal sized-groups based on their present-day galactocentric radius, for the adopted LSST magnitude limit of $g < 25$. We once again scale each CDF by $65\%$ to account for LSST's sky coverage in the Southern Hemisphere. Even for streams beyond 12~kpc, LSST will have resolving power to identify several dozen, up to potentially more than one hundred member stars per stream according to our catalog, depending on the true intrinsic radial distribution of streams in the outer halo. An extended catalog of identified GC stellar streams at large galactocentric radii will be crucial for better understanding the difference between baryonic and DM perturbations to streams, which will be key if streams are to one day probe the subhalo mass function and constrain the nature of DM.

\begin{table}[]
    \renewcommand\arraystretch{1.2}
    \centering
    \caption{The 25\%-75\% percentile range of the lowest observable stellar mass (in $M_\odot$) for streams in a given bin of galactocentric radius, averaged across streams in all four halos.}
    \begin{tabular}{l|c|c|c}
       Radial range & $0-4 \rm ~kpc$  & $4-12~\rm kpc$ & $>12 ~\rm kpc$
       \\[1mm] \hline
       \textit{Gaia} $(G<19)$  & $0.64-0.77$ & $0.35-0.70$ & $0.75-0.82$ \\
       LSST $(g < 25)$ & $0.17-0.29$ & $0.11-0.20$ & $0.22-0.59$ \\
       \textit{Roman} $(z < 27.5)$ & $0.1$ & $0.1$ & $0.1 - 0.20$ \\
    \end{tabular}
    \label{tab:massprobe}
\end{table}

Table~\ref{tab:massprobe} lists the lowest mass of stream stars detectable by the three instruments in each bin of galactocentric radius from Figure \ref{fig:gaia_radii}. For each stream in each halo, we compute the lowest mass observable according to each instrument's limiting magnitude, then for each bin we compute the 25th and 75th percentile values. The Table displays these values averaged across the four halos. Stars in the middle bin are typically at the closest heliocentric distances, which will allow LSST and \textit{Roman} to sample the mass function even below $0.2\Msun$. \textit{Gaia} is also relatively effective in this radius bin, being able to probe down to $\sim0.3\Msun$ for a quarter of the streams within this galactocentric distance range. However, in the outer bin, the \textit{Gaia} limit increases dramatically. The percentiles reported here are only for the streams in which there is more than one star below the detection limit of \textit{Gaia}; in every halo, there are streams that \textit{Gaia} misses entirely. The exact number varies according to the varying quantity and radial distribution of streams in each halo, from a minimum of 4 in 523889, to a maximum of 91 in 519311. The FIRE halo m12w, which most closely matches the MW in terms of its GC radial distribution (see Appendix \ref{Properties of Surviving GC Population}), has 19 streams for which \textit{Gaia} can observe no stars. Given the fact that at minimum several dozen stars are needed to identify a stream with current methods \citep[e.g., ][]{Malhan2018MNRASStreamfinder}, this result suggests that there are anywhere from  several dozen up to over one hundred streams in the MW halo undetectable with \textit{Gaia}.

The outer streams, at a median radius of 28.5~kpc, should still reveal stars all the way down to $0.2\Msun$ with \textit{Roman}. This means that \textit{Roman} has the potential to provide an essentially complete view of all disrupted GCs in the Milky Way.

\section{Discussion}
\label{sec:Discussion}

Our model catalog of 1471 tidal features from disrupted GCs across four cosmological halos contains a diverse set of objects including hundreds of thin streams, which can be used for many purposes such as analysis of a stream population in a MW-like galaxy, mock photometry of individual streams, modeling of streams to constrain the Galactic potential, and testing stream detection algorithms. Here we describe additional possible uses of the catalog, analyze our initial results further, and discuss possible limitations.

\subsection{Orbital Properties of Thin and Wide Streams}

\begin{figure}[!t]
    \centering
    \includegraphics[width=\columnwidth]{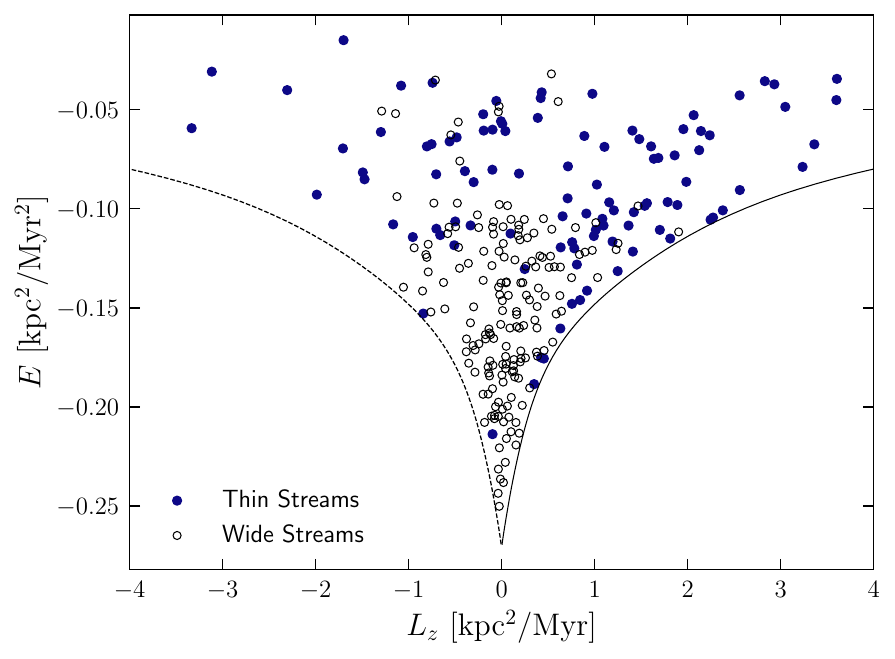}
    \caption{Characterization of orbits for all objects in 523889 in the space of the specific energy $E$ and the $z$ component of angular momentum. Tidal debris which resides deeper in the Galactic potential tend to not form thin streams. The solid line indicates the limit of a prograde circular orbit, while the dashed line represents the limit of a retrograde circular orbit.}
    \label{fig:e_lz}
\end{figure}

Given the orbital information of the GC progenitors for the last several Gyr, we can further distinguish between the types of orbits that form long, thin stellar streams and those that do not. In Figure~\ref{fig:e_lz} we show the energy-angular momentum diagram for all 287 progenitor orbits for the features in halo~523889. In our coordinate system prograde orbits are on the right (positive $L_z$) and retrograde orbits are on the left (negative $L_z$). Over 90 percent of thin streams based on our definition have energies greater than $-0.13 ~\rm kpc^2~Myr^{-2}$, indicating that orbits more strongly bound than this (thereby residing closer to the Galactic center) are heated to the point where they do not form thin coherent streams. This is also influenced by the fact that these strongly bound orbits have much smaller orbital periods, meaning they have completed more orbits within our integration time than the other clusters. While most of these strongly bound orbits form wide streams, there are also many streams on prograde circular orbits which remain thin deeper into the potential well. Notably, the current known phase-space distribution of GC streams in the MW \citep{Ibata2024ApJStreamFinderII} does not have such streams. It is possible that these streams have evaded detection thus far; they are closer to the Galactic center and appear on the sky in regions of higher background stellar density, which makes them harder to detect. It is also possible that true streams on these orbits are susceptible to heating from disk properties which are not modeled in this work. Further analysis of the orbits in connection with the resulting tidal features should help to clarify what types of streams to expect on what types of orbits.

\subsection{Impact of Future Observatories}

Our findings agree with previous results that both LSST and \textit{Roman} will greatly expand our knowledge of streams in the MW and other galaxies in the Local Group \citep{Pearson2019ApJ, Shipp2023ApJ, Aganze2024ApJ, Pearson2024}. Specifically, we find that the number of observable stars per stream will increase drastically. This provides the chance to both significantly increase the amount of confirmed members of known streams, and shows potential to find entirely new streams previously undetectable. While we have not modeled a MW background population in this work, which is a major hurdle in the detection of new streams, \cite{Pearson2024} argues that streams whose main sequence turnoff is one magnitude below the detection limit should be observable based on previous stream discoveries \citep{Shipp2018ApJ}. Applying this to our dataset, we find that every stream in the entire catalog could in principle be detected by \textit{Roman}. For LSST, we find that after accounting for its footprint on the sky that an average of 63\% of streams are detectable across the hour halos. Given that LSST can only image 65\% of the sky, this means that given this definition of detectability, it will be able to detect nearly every stream in the Southern Hemisphere.  

We can also use these data to estimate the number of streams potentially observable at present that have evaded detection. Using our adopted definition of a thin stream, there are between 60 and 124 thin streams with 100 or more observable stars with \textit{Gaia} in each halo, compared with the ~80 or so known GC stellar streams. This suggests the possibility of several dozen missing streams within the detection limit of \textit{Gaia}. However, the number may even be greater than this, since we are only tracking streams which formed in the last $3.5$~Gyr (see \citealt{Pearson2024}). One possible reason for evading detection is extinction from MW dust near the Galactic plane. Out of our set of thin streams with at least 100 stars below the \textit{Gaia} detection limit, the total number with progenitor locations on the sky with $|b| < 10^{\circ}$ ranges between 36 -- 64 for our four different halos. These streams would be partially or completely obscured in optical wavelengths due to foreground dust. A future infrared survey astrometry telescope could make it possible to detect such streams, such as the proposed GaiaNIR mission \citep{Hobbs2018IAUSGaiaNIR}.

We acknowledge that a true detailed analysis on the observability of streams for a given instrument would require modeling the background MW population and extinction from Galactic dust. There is also the matter of what data is available per instrument: for example, \textit{Gaia} also computes proper motions of potential members, which can greatly enhance the ability to detect streams even with its lower magnitude limit. However, LSST will also produce proper motions for stars over a 10-year baseline, and obtaining proper motions with \textit{Roman} is also possible in the future. Considering these facts, we predict from our findings that LSST and \textit{Roman} will expand the known number of stellar streams and give us better insight into previously known streams.

\subsection{Synthetic Photometry of Individual Streams}

\begin{figure}[!t]
    \centering
    \includegraphics[width=\columnwidth]{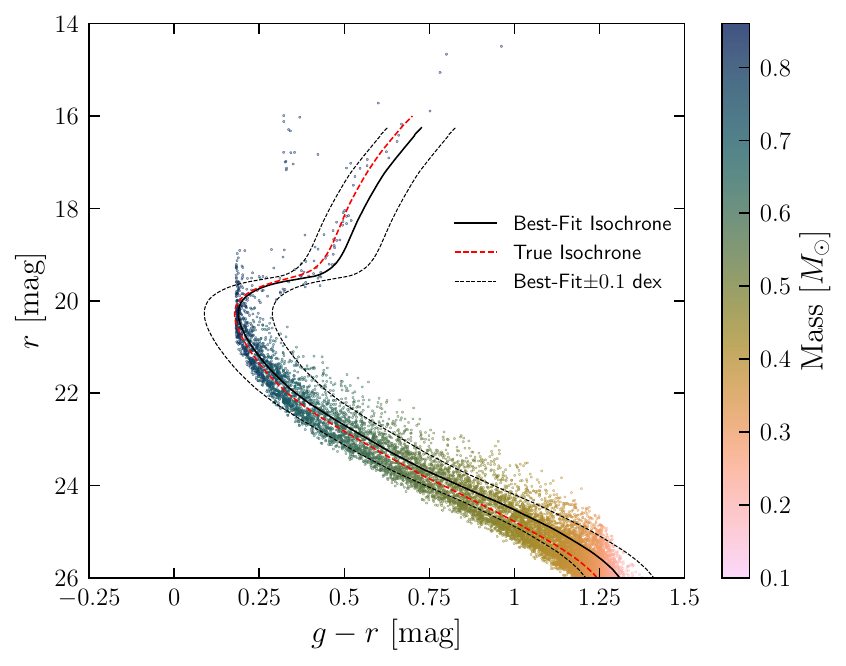}
    \caption{Color-magnitude diagram for a stream with similar apocenter and pericenter values to Pal 5, color-coded by stellar mass. The corresponding cluster in the catalog has an age of $10.36$~Gyr and we set the metallicity to $\feh = -1.56$. The MIST isochrone for a stellar population with this age and metallicity at a distance of $20.9$~kpc is shown as a dashed line. Next to it is a solid line, representing the best-fit isochrone from the data with $g < 25$ ($9.2$~Gyr, $\feh = -1.23$). The scatter of stars is due to stream stars having different distances from Earth. The dotted lines located at $\pm0.1$~dex in color from the best-fit isochrone show the bounds needed to capture 80\% of all stream members.}
    \label{fig:pal5cmd}
\end{figure}

It is also possible to analyze individual streams in the catalog and create analogs to real observed systems. An example is in Figure~\ref{fig:pal5cmd}, which shows the color-magnitude diagram (CMD) of a stream with similar mass and orbit to Pal~5. Its progenitor has an orbital pericenter of $11$~kpc, apocenter of $21$~kpc, and a present-day mass of $1.6\times10^4\Msun$. In order to match the observed properties of Pal~5, for this Figure only we keep the Sun fixed at $8.122$ kpc from the Galactic center, but perform a rotation in the $x-y$ plane to best match Pal~5's radial velocity \citep[$-58.6 ~\rm km~s^{-1}$;][]{Baumgardt2019MNRAS} and heliocentric distance \citep[$21.9$ kpc;][]{BaumgardtVasiliev2021MNRAS}. The optimal rotation gives our mock cluster a heliocentric distance of $20.9$ kpc and $v_r = -64.8 ~\rm km ~s^{-1}$. We use the MIST isochrone for an age of $10.355$ Gyr and metallicity of $\feh = -1.56$ to compute mock SDSS $g$ and $r$ magnitudes for the stars with their updated distances, assuming no extinction along the line of sight.

Stars along the stream have slightly different heliocentric distances than the progenitor, which creates measurable scatter around the original isochrone. We treat the mock CMD as if it were observed data and compute the best-fitting isochrone for all stars with $g < 25$ to the widened CMD. We sample a $100\times100$ grid of ages and metallicities ranging from $\{9,13\} ~\rm Gyr$ and $\{-2 , -0.5\} ~\rm dex$, and minimize the summed squared distance from each point to the isochrone. The best fitting isochrone from this search has an age of $9.2 ~\rm Gyr$ and $[\rm Fe/H] = -1.23$. The two curves are similar near the main sequence turnoff but begin to deviate by about 0.1 dex at $r \approx 25$, where the stars were not included in the fitting. There is also a notable discrepancy on the red giant branch where there are fewer stars. Notably, if all stars are included, we get a much closer fit giving an age of $9.97$ Gyr and $\rm [Fe/H] = -1.49$. This example shows that deeper imaging, such as that of \textit{Roman}, will allow for significantly better constraints on a stream progenitor's age and metallicity given its CMD.

We can also use this mock CMD to provide a practical guide for future observations trying to select a relatively complete sample of stream stars for follow-up spectroscopy, while minimizing contamination from the stellar background. We find that $80\%$ of the stream stars are located within $\pm 0.1$~dex in color from the best-fit isochrone for stars included in the original fit. Filtering stars based on the spread from an isochrone is commonly used in identifying stream candidates \citep[e.g.,][]{Valluri2025ApJ}, and our analysis shows that it should indeed capture the majority of the stream stars.

\subsection{Connection to Host Potential}

While not done explicitly here, the catalog along with its provided potential can be used to better understand the connection between a stellar stream population and its host galaxy. Primarily, this involves using streams to measure the enclosed mass profile of the galaxy, which can then be applied to real streams in the MW. This can also be extended to measuring the shape of the MW halo, which has been done in previous work but has produced contrasting results depending on the assumed model parameters. More work is needed along these directions, and our catalog serves as a possible useful tool.

Another promise of streams is studying the connection between the stream population and the assembly history of the host halo \citep{Carlberg2017ApJ, Bonaca2021ApJLOrbitalClustering, TingLi2022ApJ}. Our GC formation and stream models are based on cosmologically-formed halos, for which we know the true assembly history. Such a system has proved valuable for reconstruction of the galactic assembly histories with GCs \citep{YbChen2024OJAGalaxyAssembly}. With our stream catalog, a more comprehensive analysis can be applied to better understand the MW assembly history. 

\subsection{Scaling of Stream Length}

In Section~\ref{sec:Results} we found that the length of the stream population follows a power law dependence on galactocentric radius, $\sigma_{\phi_1} \propto r^{-1.8\pm0.3}$. The idea that the length distribution should follow a power law is not immediately obvious. To understand this result better, we consider an idealized model of a progenitor GC with mass $m$ on a circular orbit at radius $r$ with orbital frequency $\Omega$. We assume that the enclosed mass of the Galaxy follows the relation $M(r) \propto r^{1-\gamma}$ with $\gamma \approx 0$ at the range of radii $10-100$~kpc, where we measure the power-law relation. For the real Milky Way, \cite{Shen2022ApJ} estimate $\gamma\approx 0.4$; for our simulated galaxies, we estimate an average $\gamma\approx 0.2$. \cite{Johnston1998ApJ}, \cite{Johnston2001ApJ}, and \cite{Amorisco2015MNRAS} showed that the length of the stream is expected to grow with time as
\begin{equation}
  \sigma_{\phi_1} \sim t \, r_t \, \frac{\partial\kappa}{\partial r}  
\end{equation}
where $\kappa$ is the epicycle frequency and $r_t$ is the tidal radius. For our adopted mass distribution $\kappa = (2-\gamma)^{1/2}\, \Omega$. The instantaneous Lagrange points define the tidal radius as
\begin{equation}
  r_t \approx \left( \frac{m}{2 M(r)} \right)^{\frac{1}{3}} r
  \label{eq:rt}
\end{equation}
however for eccentric GC orbits the tidal radius may be set closer to the peri-center $r_p$; we consider this possibility below. For the circular orbit we obtain
\begin{equation}
    \sigma_{\phi_1} \propto r^{-\frac{4}{3}-\frac{\gamma}{6}} \, t
\end{equation}
and therefore, for a population of streams whose duration times are comparable, in this simple model we would expect the lengths of streams to follow a power law, albeit with a shallower slope ($\approx -1.4$) than the one found for our catalog.

One possible reason that the measured slope ($-1.8$) would be steeper is that GCs at large galactocentric radii tend to be on highly radial orbits, for which effective tidal truncation happens at smaller tidal radii than given by Equation~\ref{eq:rt}. Additionally, streams on radial orbits spend a longer duration near apocenter, where the tidal tails tend to compress as the stream stars cluster more tightly together in action space.

If we include only streams with eccentricities less than $0.4$ (of which there are 59 beyond 10 kpc) to select orbits closer to circular, we find that the best-fit slope becomes $\alpha = -1.7 \pm 0.3$, which is indeed shallower than the result with all streams but still in tension with our toy model. A more robust calculation of the length scaling would take into account non-circular orbits, as well as the expansion and compression of streams near pericenter and apocenter. We do not attempt this here, but nevertheless we find this simple model useful in trying to understand why the lengths should follow a power law relation.

\subsection{Modeling Limitations}

Selecting galaxies from cosmological simulations for our model calculation leads to two numerical limitations. First, we are limited by both the spatial and temporal resolution of the simulations. Since our model GC initial positions are assigned to particles in TNG and FIRE, they can be limited by each simulation's force-softening length. In FIRE, this is only $4$ pc for stellar particles, but for TNG it is $\sim 300$~pc, which can lead to errors in the distribution of GCs in the inner galaxy at a lookback time of $3.5$~Gyr. Moreover, the time resolution between snapshots used in the potential fitting varies between 100 and 200~Myr for TNG and 10 Myr - 1 Gyr for FIRE, which washes out important baryonic components of the potential that can impact streams, such as the spiral arms and the Galactic bar. There is also no LMC analog in our selected halos, which has been shown to leave significant perturbations on streams in its vicinity \citep[e.g., ][]{Koposov2023MNRASOCStream, Lilleengen2023MNRAS, Brooks2025ApJ}. Higher resolution large-scale simulations will be needed to model all of these affects more accurately and study their impact on the total Galactic stream population.

Second, the fitting of the potential using our BFE techniques in \texttt{agama} has its own limitations when considering the appropriate choices of $l_{\rm max}$ and $m_{\rm max}$. Selecting too low a value for these parameters, particularly $l_{\rm max}$, will smooth out too much of the halo's features and substructure. However, too high a value of $l_{\rm max}$ will cause the potential to be overfit in regions where the particle numbers are low. This is especially true close to the Galactic center, where the number of particles enclosed within a spherical shell becomes small. For example, out of all the particles present at the last snapshot for TNG halo 523889, only $3\%$ are within 1 kpc of the Galactic center, and only $11\%$ are within 4 kpc, which roughly corresponds to the scale length of the MW disk. At the same time, it is important to model the potential accurately at small galactocentric radii, since the cluster mass loss on eccentric orbits is dominated at pericenter. We made the selection of $l_{\rm max}$ and $m_{\rm max}$ after careful consideration of both competing issues (see Appendix~\ref{ap:ap}), but further improvements in this area will be essential for high-precision modeling of streams in time-evolving potentials.

In addition to these limitations, our prescription for the mass loss of GCs is further limited by the fact that Equation \ref{masslossrate} has no dependence on the internal dynamics of the cluster itself. While our model is calibrated from the N-body models of \cite{GielesGnedin2023} for GCs with typical density profiles, clusters which are more extended or more densely concentrated will have different mass-loss rates. Furthermore, recent work has shown that a population of stellar mass black holes within the GC can affect both the mass-loss rate of GCs as well as the properties of stellar streams \citep{GielesGnedin2023, Roberts2025MNRASParticleSpray, WeatherfordBonaca2025}. A mass-loss rate calculation which considers internal cluster dynamics as well as a higher resolution in the snapshot cadence of the underlying potential can more accurately capture the physics which causes gaps and density variations along stream tracks. This type of modeling will be important for understanding the differences between baryonic and dark matter perturbations along stream tracks.

\section{Conclusions}
\label{sec:conclusion}

Here, we summarize the main takeaways of this work:
\begin{itemize}
\item We extend the work of \cite{Pearson2024} by generating a catalog of GC stellar streams in four MW-like galaxies within a cosmological context of the TNG and FIRE simulations. The catalog combines the \cite{YBChen2024MNRAS} GC evolution model with the \cite{YBChen2024ParticleSpray} algorithm to generate streams in the time-evolving potentials. The catalog contains the full phase-space information for individual stars in each stream, as well as orbital history and physical properties for stream progenitors.

\item We find that the model stream length, width, and proper motions in the MW halo follow power law scalings with galactocentric radius. The best-fit slopes for the length and width are $\alpha=-1.8$ and $\alpha=-2.0$, respectively.

\item We apply mock photometry to our streams and quantify the number of observable stars per stream with \textit{Gaia}, LSST, and \textit{Roman}. We find that both LSST and \textit{Roman} may discover up to hundreds of new streams in the Milky Way. In particular, \textit{Roman} will be able to observe stars down to $0.1\Msun$ for the majority of streams.

\item We make our catalog publicly available as a tool for studying GC streams in MW-like galaxies.
\end{itemize}

The repository for the catalogs is hosted at \href{https://github.com/cholm-hansen/StreamCatalogs}{https://github.com/cholm-hansen/StreamCatalogs}, which contains a description of the catalog and its components, instructions for downloading, and an example notebook showing how to use and perform calculations with it.

\section*{Acknowledgments}
The authors thank Monica Valluri, Eric Bell, Eugene Vasiliev, Neil Ash, and Brigette Vazquez Segovia for insightful discussions. 
CHH, YC, and OYG were supported in part by National Aeronautics and Space Administration through contract NAS5-26555 for Space Telescope Science Institute programs HST-AR-16614 and JWST-GO-03433.
This research benefited from the Gravity in the Local Group conference hosted by the McWilliams Center for Cosmology and Astrophysics, Carnegie Mellon University. This research was supported in part through computational resources and services provided by Advanced Research Computing at the University of Michigan, Ann Arbor.

\paragraph{Software}
\textsc{agama} \citep{Agama2019}, \textsc{astropy} \citep{Astropy2022}, \textsc{isochrones} \citep{Isochrones2015ascl}, \textsc{matplotlib} \citep{Hunter2007matplotlib}, \textsc{numpy} \citep{Harris2020Numpy}, \textsc{scipy} \citep{Virtanen2020Scipy}

\bibliographystyle{aasjournal}

\bibliography{main}

\begin{appendix}

\section{Properties of Surviving GC Population}
\label{Properties of Surviving GC Population}

\begin{figure*}[!t]
    \centering
    \includegraphics[width=\textwidth]{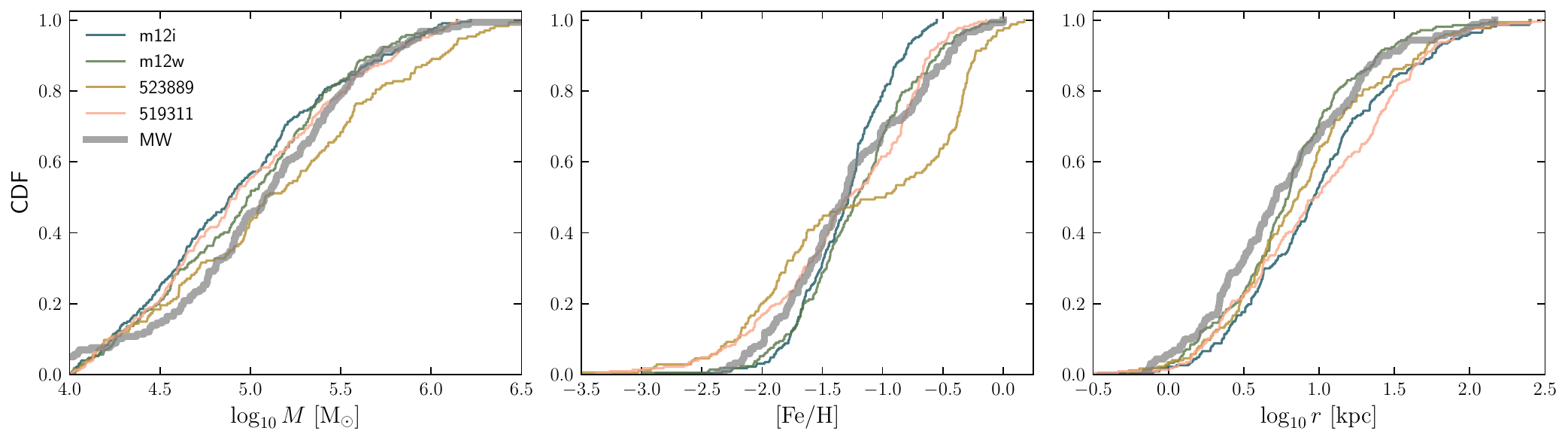}
    \caption{CDFs of the mass, metallicity, and radial distribution of surviving GCs in our catalogs at the present day. The MW is plotted for comparison. Our GCs are in good agreement with the present day mass and radial distribution of MW GCs, which are the most important parameters for generation of stellar streams.}
    \label{fig:comp_halos}
\end{figure*}

For the creation of our GC stream catalogs, we modified the GC evolution model in \cite{YBChen2024MNRAS} to output cluster masses at each snapshot and subsequently had to re-run the model on the four cosmological halos. Due to inherent random number generation in the code, our GC populations are slightly different than the published catalogs in \cite{YBChen2024MNRAS}. Here we show the properties of each GC population in comparison with the MW\footnote{The data used to calculate the MW CDFs is available at \href{https://people.smp.uq.edu.au/HolgerBaumgardt/globular/}{https://people.smp.uq.edu.au/HolgerBaumgardt/globular/}}. Figure~\ref{fig:comp_halos} shows the cumulative probability distribution functions (CDFs) for the mass, metallicity, and radial distributions of surviving GCs in our catalog, with masses greater than $10^4 \Msun$ in the present day. Observed CDFs for the MW GC population are displayed with a thicker line. Our GC populations are matched well particularly to the mass and radial distributions of the MW GCs, both of which are more important for the purposes of generating stellar streams.

\section{Calibration of the Galactic Potential}
\label{ap:ap}

\begin{figure}[!t]
    \centering
    \includegraphics[width=\columnwidth]{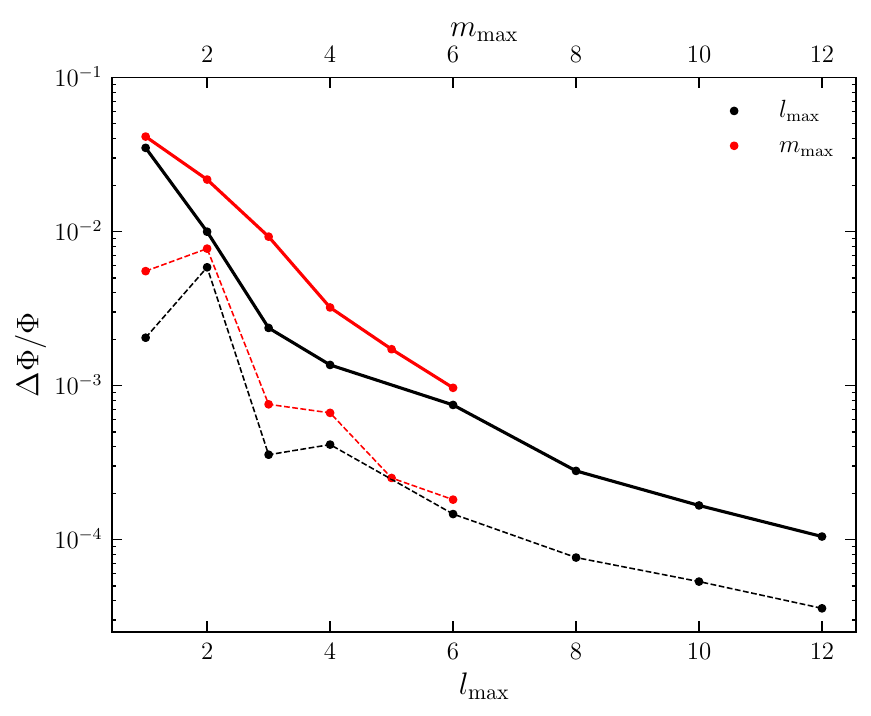}
    \caption{Convergence of the halo and disk potentials for TNG galaxy 523889. For each data point, we increment $l_{\rm max}$ or $m_{\rm max}$ and plot the average absolute relative difference in the value of the potential across the initial cluster locations compared to the previous point. The solid lines show the convergence using the first TNG snapshot used in our model and the dotted lines shows the last snapshot.}
    \label{fig:multipole_convergence}
\end{figure}

\begin{figure}[!t]
    \centering
    \includegraphics[width=\columnwidth]{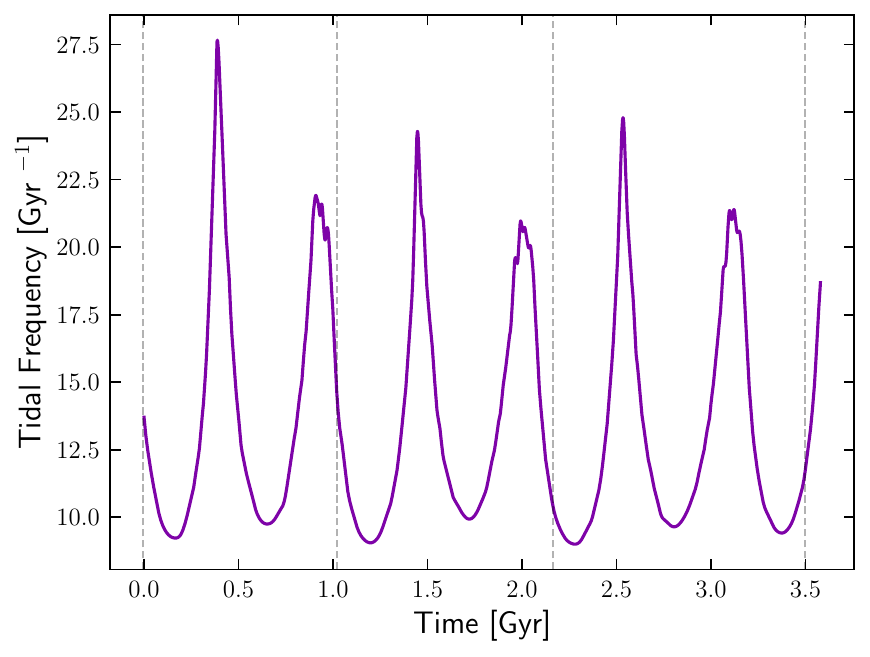}
    \caption{An example of the effective tidal frequency (Eq.~\ref{eq:omegatid}) along the orbit of a model cluster with similar apocenter and pericenter to Pal 5. The line shows the analytic calculation of the tidal tensor using the BFE potential in \texttt{agama}. Each spike corresponds to a tidal shock near the orbital pericenter. The vertical dashed lines indicate the epochs when full TNG snapshots are available to calculate the tidal tensor directly from the baseline simulation. The only four available data points emphasize the advantage of using the \texttt{agama} orbits to compute $\Omegatid$.}
    \label{fig:agamatngcomp}
\end{figure}

From a lookback time of $\sim$3.5 Gyr until the present, we use potentials fit with BFEs in \texttt{agama} to calculate orbits of progenitor GCs and tidally stripped stars, as well as for the calculation of the mass-loss rate of the GCs along their orbit. A two-component BFE is fit to simulation snapshots, which is then linearly interpolated in time to create the time-dependent potential. The DM particles are fit with a multipole expansion:
\begin{equation}
    \Phi(r,\theta,\phi) = \sum_{l=0}^{l_{\mathrm{max}}} \sum_{m=-l}^{l} \Phi_{l,m}(r) Y_l^m(\theta,\phi)
\end{equation}
where $Y_l^m(\theta,\phi)$ are the usual spherical harmonics. The star and gas particles are modeled with an azimuthal Fourier harmonic expansion in $\phi$:
\begin{equation}
    \Phi(R,z,\phi) = \sum_{m=0}^{m_{\mathrm{max}}}\Phi_m(R,z)e^{im\phi}
\end{equation}

Before the potentials are fit, we rotate the particles in each snapshot to a common reference frame aligned with the principal axis of the disk particles at the last snapshot. This is needed for proper fitting of the Fourier expansion terms, in addition to needing one consistent coordinate system in order to interpolate between adjacent snapshots. While the disk orientation does change slightly between the first and last snapshot, we find that for three of the four halos used, this change is only a few degrees, which does not have a significant impact on the fit. For the one with a significant change in the disk orientation (m12i), we opt for a BFE using only multipole expansion, as described in Section~\ref{sec:simulation}.

For both the halo and disk potentials, we sample coefficients $\Phi_{l,m}(r)$ and $\Phi_{m}(R,z)$ on logarithmically spaced grids and spline interpolate between the individual points to get the final analytic potential at each snapshot. The time-dependent potential is then constructed by linearly interpolating between the individual snapshot potentials. For the multipole expansion, no symmetry is assumed and all $m$ terms are kept for each corresponding $l$. To determine the highest order of $l$ in the multipole expansion and $m$ in the Fourier harmonic expansion, we perform a convergence test by evaluating the change in each respective potential at the locations of the GCs 3.5 Gyr ago for one of our halos (523889). For the disk potential, we start at $m_{\rm max}=1$ and compute the relative absolute change in potential between $m_{\rm max} - 1$ and $m_{\rm max}$, averaged at the locations of all the GCs. The same is done for the halo potential, except we increment $l_{\max}$ by 2 instead of 1. The result is shown in Figure~\ref{fig:multipole_convergence}. The shorter lines are for $m_{\rm max}$ and the longer ones are for $l_{\rm max}$. Additionally, the solid lines are for the first snapshot used for the BFEs ($z=0.3$), and the dashed lines are for the last snapshot ($z=0$). Below $\Delta \Phi / \Phi$ of $10^{-3}$, the error in the orbit will be dominated by the resolution of the snapshot interpolation rather than the fits to the potentials themselves. Based on the convergence of both the disk and halo potentials, we adopt $l_{\rm max} = 12$ and $m_{\rm max} = 6$.

\section{Reconstructed Orbits}
\label{Reconstructed Orbits}
\begin{figure*}[!t]
    \centering
    \includegraphics[width=\textwidth]{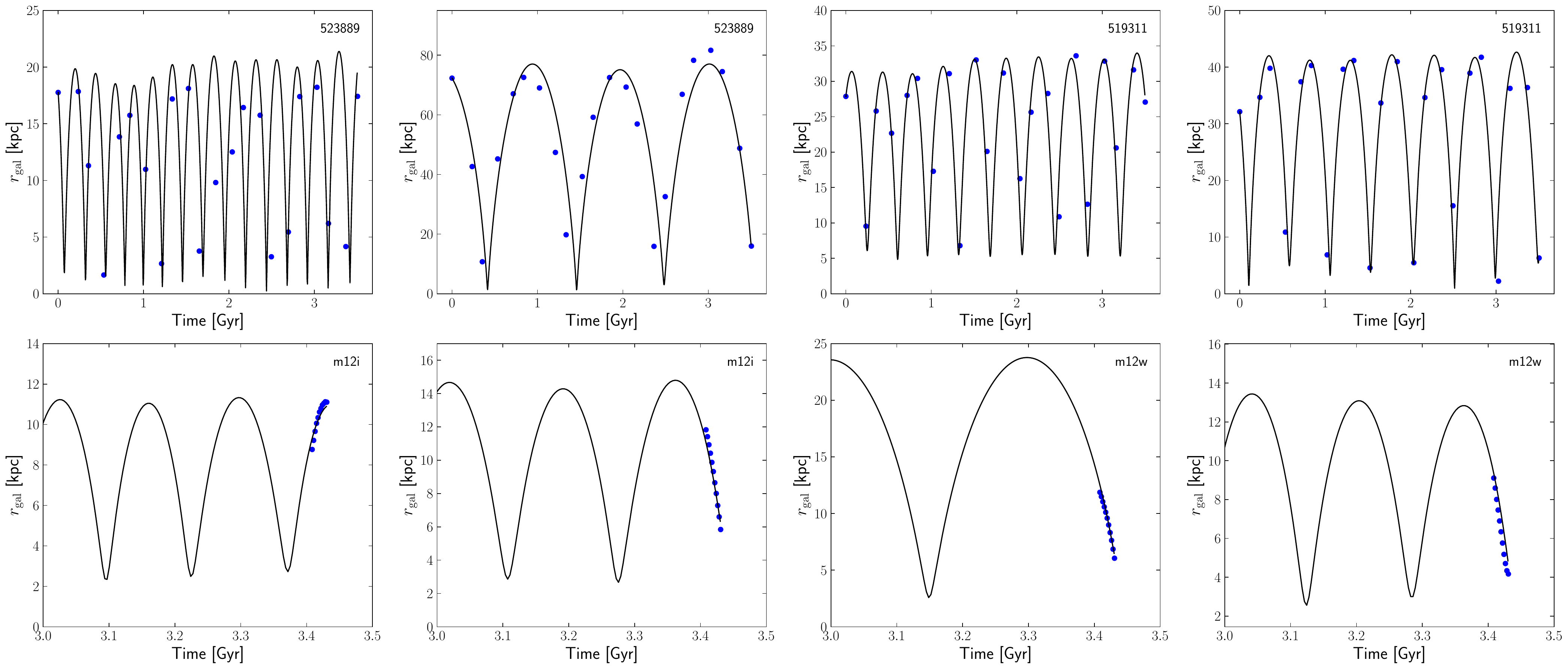}
    \caption{Reconstructed orbits from our method outlined in Section \ref{sec:model}. The blue points represent the true orbit from the cosmological simulation; the black line is our reconstructed orbit calculated from \texttt{agama} using our BFE potentials. The background cosmological simulation for each particular orbit is shown in the top right of each panel.}
    \label{fig:reconstructedorbits}
\end{figure*}

In Figure \ref{fig:reconstructedorbits}, we show some reconstructed orbits from our fitted potentials compared to the true particle orbit in the respective cosmological simulation. For each individual halo we show two reconstructed GC orbits for a total of eight. In each panel, the blue dots show the galactocentric radius of the GC (more specifically, the galactocentric radius of the particle in the simulation we assign to represent the GC) at each snapshot used in this work. The top row shows four orbits from the two TNG50 galaxies, and the bottom row shows orbits from the FIRE galaxies. For FIRE, we opt to show the last $\sim 500$ Myr of the orbit, which contains 10 snapshots with high resolution. For TNG50 we show the full integrated orbit used in this work, approximately $3.5$ Gyr for each halo. Our method reliably recovers the pericenter and apocenter values for most orbits.

\section{Comparison of Low and High Time Resolution for Mass-Loss Calculation}
\label{TidalFreqAppendix}

The fitting of the BFE potentials allows us to calculate the mass-loss rate along the GC orbits at higher resolution than is available with the GC evolution model in \cite{YBChen2024MNRAS}. In Figure \ref{fig:agamatngcomp}, we show our computed effective tidal frequency (defined from the eigenvalues of the tidal tensor), for a selected cluster in the catalog on a similar orbit to Pal~5. Each peak in $\Omega_{\rm tid}$ is from a pericenter passage of the GC on its orbit. By comparison, the \cite{YBChen2024MNRAS} model only computes $\Omega_{\rm tid}$ at full TNG snapshots, the timestamps of which are shown with vertical dashed lines. For our higher resolution calculation using BFE, \texttt{agama} interpolates the computed solution to give us the values of $\Omega_{\rm tid}$ on $10^3$ uniformly space intervals within the last $\sim$3.5 Gyr. Our higher resolution calculation is used to better capture the mass-loss of the GCs along their orbits.

\section{Comparison with N-body}
\label{NbodyComp}

Here, we perform an N-body realization of one stream from our catalog to assess the validity of the \cite{YBChen2024ParticleSpray} algorithm in a time-evolving, asymmetric halo. For our N-body test we use the multipole gravity solver \texttt{falcON} \citep{Dehen2000, Dehen2002falcon} with our TNG potential from halo 523889 as the external field. We select stream 29 from the 523889 catalog as our test case, which has a pericenter of $\sim$11~kpc and an apocenter of $\sim$21~kpc, similar to the stellar stream Pal 5. We initialize the cluster as a Plummer profile with a scale radius of $4$ pc and initial mass $2.4\times 10^4 \Msun$, which are the same initial conditions provided to the Plummer potential in the catalog realization of the stream. We take the particle masses to be the mean particle mass of stars in the catalog stream, which in this case is $0.33\Msun$, making the total number of particles $7.3\times 10^4$. We place the N-body cluster at the same initial location in the galaxy and evolve it for $3.5$ Gyr. Based on the optimal softening length relation from \cite{Athanassoula2000MNRASsoftening}, we take $\varepsilon =0.2$ pc and use a timestep of $2^{-13} ~\mathrm{km ~s^{-1}~kpc^{-1}} \approx 0.1 ~\mathrm{Myr}$.

\begin{figure*}[!t]
    \centering
    \includegraphics[width=\textwidth]{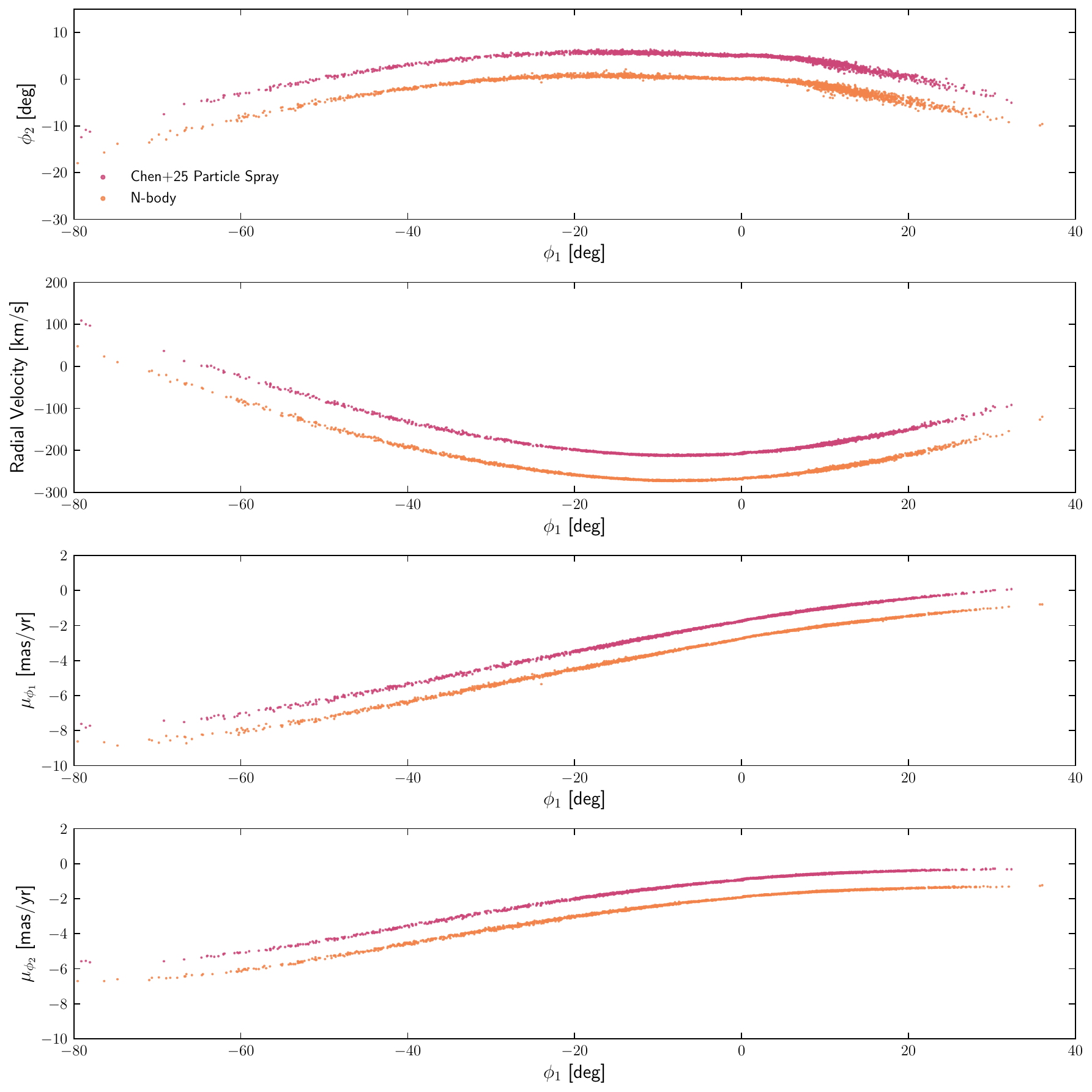}
    \caption{N-body comparison of one stream from the catalog. In each panel, the top stream is generated from the \cite{YBChen2024ParticleSpray} particle-spray algorithm, while the bottom stream is generated using \texttt{falcON}. For clarity of display, the streams are offset by $5^{\circ}, 60~\mathrm{km~s^{-1}}$, and $1~\mathrm{mas~yr^{-1}}$, respectively. The streams are nearly indistinguishable in their distributions along $\phi_1$, indicating that the \cite{YBChen2024ParticleSpray} algorithm is a valid approximation for full N-body integration even in a time-evolving, non-axisymmetric halo.}
    \label{fig:nbodycomp}
\end{figure*}

Figure~\ref{fig:nbodycomp} compares the positions, proper motions, and radial velocity of the stream generated with \texttt{falcON} compared to the catalog stream generated with particle-spray. The progenitor locations for each stream are approximately located at $\phi_1 \approx 0^{\circ}$. For plotting purposes, we do not show the progenitor from the N-body stream since we do not resolve the progenitor in the particle-spray version. Additionally, the particle-spray stream has more particles than the final N-body stream, so we randomly down-sample the particle-spray stream to show the same number of particles in both streams. Finally, we apply offsets of $(5^{\circ}, 60~\mathrm{km~s^{-1}},1~\mathrm{mas~yr^{-1}}, 1~\mathrm{mas~yr^{-1}})$ to the particle-spray stream on the y-axis so that the two streams can be visually compared. Comparison of the tracks along $\phi_1$ shows that the streams are nearly indistinguishable in morphology, proper motions, and radial velocity, with good agreement for both the overall stream length and width. Using the same classification statistics from Figure \ref{fig:phi1phi2plots}, the percent differences between the particle-spray stream and N-body stream $(\sigma_{\phi_1}, \sigma_{\phi_2}, \sigma_{\mu_{\phi_1}}, \sigma_{\mu_{\phi_2}})$ are $(-13\%, -26\%, -12\%, -12\%)$. The difference in the radial velocity spread, $\sigma_{rv}$ is $-6\%$. 

Across these metrics, the particle-spray model tends to have a slightly smaller dispersion than the N-body model. In addition to inherent errors caused by differences between the two methods (N-body versus particle-spray), our prescription for GC mass loss may also contribute to the difference in some quantities. For example, part of the difference in $\sigma_{\phi_1}$ could be attributed to the number of stars unbound per unit time according to Equation \ref{masslossrate} versus the value from direct N-body. We also find that different choices for the softening length have an effect on how well the resulting stream matches the particle-spray; for example, a trial run with the same setup but a softening length of $\varepsilon = 1~\mathrm{pc}$ produces better agreement with the particle-spray model, with percent differences in $(\sigma_{\phi_1,} \sigma_{\phi_2}, \sigma_{\mu_{\phi_1}}, \sigma_{\mu_{\phi_2}}, \sigma_{rv})$ being (-6\%, -17\%, -3\%, -1\%, +9\%). However, given the size of our initial cluster and the number of particles used, this softening length is somewhat large to accurately capture the internal cluster dynamics.

Aside from the width, the deviation of all other metrics are still around $10\%$, which is the average value quoted in \cite{YBChen2024ParticleSpray} for the potential models tested in their work. Based on this as well as the visual similarity of the stream tracks, we conclude that the \cite{YBChen2024ParticleSpray} algorithm is still a valid approximation for modeling cold streams in a time-varying, non-axisymmetric halo.

~\\
\end{appendix}
\end{document}